\documentclass[aps,prd,a4paper,twocolumn,showpacs,nofootinbib]{revtex4}
\usepackage{amssymb,amsmath,latexsym,mathrsfs}
\usepackage{graphicx}
\usepackage{epsfig}

\def\dj{\hbox{d\kern-0.347em \vrule width 0.3em height 1.252ex depth
-1.21ex \kern 0.051em}}

\def\ran{\rangle}
\def\lan{\langle}

\def\in{{\rm{in}}}

\newcommand{\be}{\begin{equation}}
\newcommand{\ee}{\end{equation}}
\newcommand{\ben}{\begin{equation*}}
\newcommand{\een}{\end{equation*}}
\newcommand{\bea}{\begin{eqnarray}}
\newcommand{\eea}{\end{eqnarray}}
\newcommand{\bean}{\begin{eqnarray*}}
\newcommand{\eean}{\end{eqnarray*}}
\newcommand{\brr}{\begin{array}}
\newcommand{\err}{\end{array}}
\newcommand{\bc}{\begin{center}}
\newcommand{\ec}{\end{center}}

\newcommand{\eg}{\mbox{\it e.g.~}}
\newcommand{\ie}{\mbox{\it i.e.~}}

\newcommand{\lsim}{\,\raisebox{-0.6ex}{$\buildrel < \over \sim$}\,}

\newcommand{\bk}{{\mathbf k}}

\newcommand{\bp}{{\mathbf p}}

\newcommand{\bB}{{\mathbf B}}

\newcommand{\bx}{{\mathbf x}}

\newcommand{\bq}{{\mathbf q}}

\newcommand{\BB}{{\cal B}}
\newcommand{\GG}{{\cal G}}
\newcommand{\HH}{{\cal H}}
\newcommand{\MM}{{\cal M}}
\newcommand{\PP}{{\cal P}}
\newcommand{\RR}{{\cal R}}

\newcommand{\cd}{\cdot}
\newcommand{\al}{\alpha}

\newcommand{\de}{\delta}

\newcommand{\ka}{\kappa}

\newcommand{\la}{\lambda}
\newcommand{\Om}{\Omega}

\newcommand{\si}{\sigma}
\newcommand{\Si}{\Sigma}

\newcommand{\dd}{\partial}
\newcommand{\mr}{\mathrm}

\begin{document}

\title{Interactions of cosmological gravitational waves and magnetic fields}
\author{Elisa Fenu}
\email{elisa.fenu@unige.ch}
\author{Ruth Durrer}
\email{ruth.durrer@unige.ch}
\affiliation{D\'epartement de Physique Th\'eorique, Universit\'e de
Gen\`eve, 24 quai Ernest Ansermet, 1211 Gen\`eve 4, Switzerland.}

\date{\today}

\begin{abstract}
The energy momentum tensor of a magnetic field always contains a spin-2
component in its anisotropic stress and therefore generates gravitational
waves. It has been argued in the literature (Caprini \& Durrer~\cite{CD})
that this gravitational wave
production can be very strong and that back-reaction cannot be
neglected. On the other hand, a gravitational wave background does affect
the evolution of magnetic fields. It has also been argued
(Tsagas et al.~\cite{Tsagas:2001ak},~\cite{Tsagas:2005ki})
that this can lead to a very strong amplification
of a primordial magnetic field. In this paper we revisit these claims
and study back reaction to second order.
\end{abstract}


\maketitle

\section{Introduction}
Wherever we can measure them in the Universe, magnetic fields of
$0.5$ to several micro Gauss are present. They have been found in ordinary 
galaxies~\cite{obs1} like
ours, but also in galaxies at relatively high redshift~\cite{obs2} and in galaxy
clusters~\cite{obs3}. It is still unknown where these cosmological
magnetic fields
come from. Are they primordial, \ie~ generated in the early universe~\cite{nl},
or did they form later on by some non-linear aspect of structure
formation, like the Harrison mechanism which works once vorticity
or, more generically, turbulence has developed~\cite{Har}?

In addition, once
initial fields are generated, it is still unclear whether
they are strongly amplified by a dynamo mechanism or only moderately
by contraction. Since the cosmic plasma is an excellent conductor, the
magneto-hydrodynamic (MHD) approximation can be employed which implies that
the magnetic field lines are frozen in during structure formation. Therefore,
as long as non-linear magnetic field generation can be neglected, the
magnetic field scales inversely proportional to the area, so that
$B/\rho^{2/3}$ is roughly constant during structure formation. Here $\rho$ is
the energy (or matter) density of the cosmic plasma. For galaxies,
with a density of about $\rho_{\rm gal} \sim 10^5\bar\rho$ this means that
simple contraction will enhance magnetic fields by approximately $10^3$,
$\bar\rho$ is the mean density.
Hence, if no dynamo is active during galaxy formation, initial fields of
$B_{\rm in} \sim 10^{-9}$Gauss are needed. On the other hand, non-linear
dynamo action can enhance the magnetic field exponentially by a factor up to
$10^{15}$, so that initial fields as tiny as $B_{\rm in} \sim 10^{-21}$Gauss might
suffice~\cite{axel}. However, since this enhancement is exponentially
sensitive to the age of the Universe, it remains unclear how it can generate
the magnetic fields in galaxies at redshifts of $z \sim 1$ or more, where the
age of the Universe was at most half its present value reducing the
amplification factor to less than $10^8$.

Another problem of cosmic magnetic fields is that primordial generation of
fields usually leads to a very blue magnetic field energy spectrum,
\be  \frac{{\rm d}\rho_B}{{\rm d}\log k} \propto k^{M+3}~, \ee
where $M=2$ for ``causally'' produced magnetic fields~\cite{DC} and
$M\sim 0$ for typical inflationary production mechanisms~\cite{WT}. Such
blue magnetic field spectra are strongly constrained by their gravity wave
production~\cite{CD} and cannot lead to the large scale fields observed
today. The only solution to the problem might lie in an "inverse cascade"
by which energy is transferred from small to larger scales. Since within
the linearized approximation each Fourier mode evolves independently,
such a cascade is inherently non-linear. Within standard MHD is has been
shown that only helical magnetic fields can lead to inverse
cascade~\cite{axel}.

In this work, we want to address a weakly non-linear effect which has not
been considered in~\cite{axel}, namely the interaction of gravitational
waves and magnetic fields. We shall study how this interaction can modify
the magnetic field spectrum. We also re-interpret a finding
by Tsagas~\cite{Tsagas:2005ki}, where the interaction between gravitational
waves and magnetic fields has been interpreted as "resonant amplification".
Similar conclusions are drawn in Refs.~\cite{D1},~\cite{D2}.
However, in this last article it is
noted that the amplification can take place only on super-horizon scales.
And even though
Ref.~\cite{D2} does mention that there is no amplification in the 
long-wavelength limit, they do not really quantify this statement.

We show that the build up of magnetic fields due to their interaction with
gravitational waves is at most logarithmic and thus comparable to the
generation of gravitational waves by magnetic fields.

Furthermore, in Ref.~\cite{D2} it has also been 
pointed out that the super-horizon "amplification"
is independent of the fact whether the plasma is highly conducting or not.
This seems physically reasonable as currents generated by electromagnetic 
fields can act only causally, \ie on sub-horizon scales. 
An animated discussion on this subject follwed the above publications and 
can be found in Refs.~\cite{Tsagas:2005qx},~\cite{Betschart:2007ts}. 
Here the role of a finite conductivity in an expanding Universe 
is addressed but controversal final conclusions have been reached.

The main advantage of our treatment is that we express the 
relevant results entirely in terms of physical, measurable quantities, which
renders the interpretation straight forward. We actually find for the 
density parameters of second order perturbations that, once the scales 
considered are inside the horizon,
\bea
\Om_B^{(2)} &\simeq& \Om_B^{(1)}\Om_{\mr{GW}}^{(1)}  \simeq 
\Om_B^{(1)}\left(\frac{H_{\mr{inf}}}{M_{\rm P}}\right)^2~, \label{iB2}\\
\Om_{\mr{GW}}^{(2)} &\simeq& \left[\Om_B^{(1)}\right]^2 + \left[
        \Om_{\mr{GW}}^{(1)}\right]^2 ~,
\eea
as one probably would expect naively.
Even though most parts of this result can already be found in the above cited
papers, they are interpreted there in a different way, and especially in
Eq.~(\ref{iB2}) it is not always noted that the factor $\Om_{\mr{GW}}^{(1)}$ 
always has to remain small. 

The paper is organized as follows. In the next section we set up the fully
non-linear equations for the evolution of magnetic fields in the
relativistic MHD approximation. We use the 3+1 formalism and closely follow
the derivation given in Ref.~\cite{Barrow:2006ch}. Since we are mainly
interested in gravitational waves, we specialize to the vorticity free case.
In Section~III we consider linear perturbations. We solve the linear
perturbation equations for gravitational waves and magnetic fields
with given initial conditions. We also derive the evolution of the
corresponding energy densities. This part is not new and mainly included for
completeness and to fix the notation for the subsequent Section~IV, where we
solve the second order equations. On this level the gravitational waves
interact with the magnetic field. We calculate the second order magnetic
field generated by this interaction and show that for reasonable values for
the first order perturbations, its energy density remains always much
smaller than the energy density of the first order contributions. In this
sense, one cannot speak of resonant amplification. In Section~V
we summarize our results and draw some conclusions.

Throughout this work we use the metric signature $(-,+,+,+)$. Conformal time
is denoted by $t$ and we neglect the background curvature of the Universe,
$K=0$. Spacetime indices are denoted by lower case Greek letters, $\mu,\nu$,
while lower case Latin letters, $i,j$ are used for spatial indices. Most of our
calculations are performed in the radiation dominated era and we shall often
use the expression
$$ a(t)= H_\in a_\in^2 t $$
for the scale factor.

\section{The basic equations}
We work in the MHD approximation, where we assume high
conductivity. The electric field is then small compared to the 
magnetic field in the baryon
rest-frame which we take to be the frame of our ``fundamental observer''.
In addition, we assume the velocity $u^\mu$ of this fundamental observer to be
vorticity-free and we neglect acceleration. According to Frobenius' theorem
$u$ is  hyper-surface orthogonal and we can choose spatial
coordinates in the three-space orthogonal to $u$.
Furthermore, in the early Universe which is of interest to us, the
dominant radiation and baryons are tightly coupled so that the energy
flux is also given by $u$ and we can set the heat flux $q=0$.
In our vorticity free frame, the
magnetic part of the Weyl tensor, $H_{ij}$ is related to the shear
simply by
$$
H_{ij} = \mr{curl}\si_{ij} ~,
$$
where $\mr{curl}$ is the 3-dim curl on the hyper-surface normal to $u$.
Here $\si$ is the shear of $u$ given by
\bean
\si_{\mu\nu} &\equiv& \frac{1}{2}\left(u_{\mu;\nu} + u_{\mu;\nu}\right)
  -\frac{1}{3}\Theta \tilde{p}_{\mu\nu} ~,\\
\Theta  &\equiv& u^\mu_{;\mu}  ~\mbox{ and}\\
 \tilde{p}_{\mu\nu}  &\equiv& g_{\mu\nu} + u_\mu u_\nu ~.
\eean
Note that the normalization of $u$ implies
   $0=u^\nu u_{\nu;\mu} \propto u_{0;\mu}$.
The gravito-magnetic interaction can then be described by the
following equations, see~\cite{Barrow:2006ch}
\begin{eqnarray}
     \nabla_u{E}_{ij}&=&-\Theta E_{ij}-\frac{1}{2}\kappa\left
      (\rho+p+\frac{1}{6\pi}B^2
      \right)\sigma_{ij} \nonumber \\  && -D^2\sigma_{ij}-
      \kappa\frac{1}{2}\nabla_u{\Pi}_{ij}   -\frac{1}{6}\Theta
      \kappa\Pi_{ij} \nonumber \\      &&
      + 3{\sigma_{\langle i}}^n E_{j \rangle n}
      -\frac{1}{2}\kappa{\sigma_{\langle i}}^n
      \Pi_{j \rangle n} \,,
      \label{E}  \\
    \nabla_u{\sigma}_{ij}&=& -E_{ij} +\frac{1}{2}
      \kappa\Pi_{ij} \! - \!
      {\sigma_{\langle i}}^n \sigma_{j \rangle n}-
      \frac{2}{3}\Theta\sigma_{ij} \,,
      \label{shear}  \\
    \nabla_u{B}_i &=& -\frac{2}{3} \tilde{p}_{ij}\Theta B^j +
      \sigma_{ij}B^j \,,
      \label{B}  \\
    \nabla_u{\Theta}&=&-\frac{1}{3} \Theta^2 \! -\frac{1}{2}\kappa\!
        \left(\! \rho\! +\! 3p\! +\! \frac{1}{4\pi}B^2\! \right) \!
        - \! 2\sigma^2\,.
       \label{theta}
  \end{eqnarray}
Here $E_{ij}$ is the electric part of the Weyl tensor, $\rho$ and
$p$ are the energy density and pressure of the cosmic fluid which is
assumed to follow the motion of the baryons (like, \eg radiation
before decoupling), $\ka=8\pi G$ is the gravitational coupling constant
and $B_i$ is the magnetic field. 
We have neglected the electric field in the above equations, 
since we assumed it to be much smaller that the magnetic
field, \ie $B^2 \gg E^2$.
The covariant derivative in direction
$u$ is denoted by $\nabla_u$ and the brackets indicate symmetrization and
trace subtraction,
$$ X_{\lan ij\ran} = \frac{1}{2}\left(X_{ij} + X_{ji}\right)
  -\frac{1}{3}\tilde{p}_{ij}{X_m}^m ~. $$
$D^2$ is the Laplace operator on the hyper-surface orthogonal to $u$.
The scalars $B^2$ and $\si^2$ are simply $\sigma^2 \equiv \sigma_{ij}
\sigma^{ij} /2 $ and $B^2 \equiv B_{i} B^{i}$.

In addition to this we have the Einstein equation, the spatial part
of which yields
\bea
   \mathcal{R}_{ij} &=& E_{ij}+\frac{2}{3}\left(\kappa\rho
        +\frac{1}{8\pi}\kappa
        B^2-\frac{1}{3}\Theta^2+\sigma^2 \right)\tilde{p}_{ij}
    \nonumber \\ && +\frac{1}{2}
        \kappa\Pi_{ij}-\frac{1}{3}\Theta\sigma_{ij}
        +\sigma_{n\langle i} {\sigma^{n}}_{j \rangle} ~,
\eea
and its trace, the generalized Friedmann equation,
\be
\frac{1}{3}\Theta^2 + \frac{1}{2}\RR = \kappa\rho
        +\frac{1}{8\pi}\kappa
        B^2+\sigma^2~.
\ee
Here $\RR_{ij}$ is the Ricci tensor on the spatial hyper-surface and
$\RR$ is its trace.

From this system we can derive  second order equations for
$\si_{ij}$ and $B_i$ which are
\begin{eqnarray}
 &&  \nabla_u \nabla_u{\sigma}_{ij} -  D^2\sigma_{ij}
       +\frac{5}{3}\Theta\nabla_u{\sigma}_{ij}  \nonumber \\ &&
+ \left( \frac{4}{9}\Theta^2
       -\frac{3}{2}\kappa p-\frac{5}{6}\kappa \rho-
   \frac{1}{6\pi}\kappa  B^2 -\frac{4}{3}\sigma^2 \right)\sigma_{ij} =
       \nonumber \\ && \qquad
       \kappa \nabla_u{\Pi}_{ij}+
       \frac{2}{3}\Theta \kappa \Pi_{ij}+
       \frac{2}{3}\kappa B^2\sigma_{ij}+
       \Theta{\sigma_{\langle i}}^{n} \sigma_{j\rangle n}
    \nonumber \\ && \qquad
      +2{\sigma_{\langle i}}^{n} \nabla_u{\sigma}_{j\rangle n}-
       {\nabla_u{\sigma}_{\langle i}}^{n}\sigma_{j\rangle n}
       - \kappa {\sigma_{\langle i}}^{n} \Pi_{j\rangle n}
       \nonumber \\  && \qquad
      + \frac{1}{3}\sigma^2\sigma_{ij}+
       3{\sigma_{\langle i}}^{n}
       \left[\frac{1}{2}\left({\sigma_{j\rangle}}^m \sigma_{nm}
           +{\sigma_n}^m \sigma_{j\rangle m}\right) \right.
            \nonumber \\  && \qquad
         -\frac{2}{3}\delta_{j\rangle n} \sigma^2 \bigg]   \;,
       \label{eq. shear}
\end{eqnarray}
and
\begin{eqnarray}
&&   \nabla_u \nabla_u{B}_{i} - D^2 B_{i}
       +\frac{5}{3}\Theta \nabla_uB_{i}  \nonumber \\  &&
+ \left(\frac{1}{3}\kappa \rho -\kappa p
       + \frac{2}{9}\Theta^2+ \frac{1}{12\pi}\kappa B^2
       + \frac{2}{3}\sigma^2 \right)B_{i}=
       \nonumber \\  && \qquad
       \sigma_{ij}\nabla_u{B}^{j}
       +2\Theta\sigma_{ij} B^{j}+
       2(\nabla_u{\sigma}_{ij}) B^{j}     \nonumber \\  && \qquad
       -\frac{3}{2} \kappa \Pi_{ij} B^{j}+
       \sigma_{\langle i}^n \sigma_{j \rangle n} B^j  
       + {\rm curl}J_i \;.
       \label{eq. B}
  \end{eqnarray}
Eq.~(\ref{eq. B}) can be obtained from  Eq.~(40) of
\cite{Tsagas:2004kv} when setting $A_i =0$, $\omega_{ij} =0$ and
$q_i =0$. $J_i$ stands for the 3-dimensional
current. Eq.~(\ref{eq. B}) is
obtained without neglecting the electric field. The term curl$E_i$, which is 
present in the original Maxwell equation which reduces to 
Eq.~(\ref{B}) if $E_i=0$~\cite{Barrow:2006ch}
results in the Laplacian term $D^2 B_i$ and  
terms proportional to the wavenumber $k$ times the electric field [see 
Eq.~(40) of \cite{Tsagas:2004kv}]. We have neglected these latter 
contributions in the above equation, since they
are only relevant inside the horizon ($kt \gg 1$), where we can neglect 
the source term of the equations, as we shall argue in the following.

In a regime of low conductivity we can neglect also the current in
Eq.~(\ref{eq. B}) and the magnetic field obeys to the above wave 
equation, while in a very hight conductivity case we should directly set the
electric field $E_i=0$ from the beginning and solve Eq.~(\ref{B}), obtaining a 
power-low behaviour with respect to time for $B_i$.
In both cases we find that the behaviour in time of the induced second order
magnetic field $B_i^{(2)}$ is the same on super-horizon scales (up to 
uncertain logarithmic corrections). We interpret this as the insensitivity  
of super-horizon perturbations to plasma properties like conductivity.

Inside the horizon, we neglect  the source term. This is motivated by 
the Green function of the damped wave equation obtained when 
linearizing~(\ref{eq. B}), which rapidly oscillates on
sub-horizon scales.
For  Eq.~(\ref{B}) it is not the Green function but the source term
$\sigma_{ij}^{(1)} B_{(1)}^j$ which oscillates when $kt \gg 1$, since 
gravity waves start oscillating at horizon crossing. Therefore again, 
the sub-horizon amplification is unimportant. The same conclusion is actually
drawn in Ref.~\cite{D2}, where the fluid velocities are not neglected.

In the following we shall consider these equations in first and second
perturbative orders with respect to a spatially flat Friedmann background,
$$ {\rm d}s^2 = a^2(-{\rm d}t^2+\de_{ij}{\rm d}x^i {\rm d}x^j) \,.$$
We neglect a possible spatial curvature of the background and work
with conformal time $t$. The time dependence of the scale factor $a$ is
determined by the Friedmann equation,
\bean
 && \left(\frac{\dot a}{a}\right)^2 = \frac{\ka}{3}\rho a^2 \quad\mbox{and}\\
 && \dot\rho = -3(1+w)\rho \left(\frac{\dot a}{a}\right) ~, \quad w =p/\rho ~.
\eean

\section{First order perturbations}
\subsection{Magnetic fields}
A background Friedmann Universe can of course not contain a magnetic
field since the latter always generates anisotropic stresses
$\Pi_{ij}\neq 0$ which break isotropy. When considering a magnetic
field as a first order perturbation, Eq.~(\ref{B}) leads in first order to
\be
\dot B^{(1)}_i = -\frac{\dot a}{a}B^{(1)}_i   \;.
\ee
For this we use that to lowest order $u=a^{-1}\dd_t$ and $(\nabla_u
B)_i =a^{-1}(\dd_t -\dot a/a )B_i$. Furthermore
$\tilde{p}_{ij}=g_{ij}=a^2\de_{ij}$ and $\Theta =3 \dot a/a^2$.
This is solved by
\bea
 B^{(1)}_i(\bx,t) &=& B^{(1)}_{i\,\in}(\bx)\frac{a_{\in}}{a(t)}~,
 \nonumber \\
 B^{i(1)}(\bx,t) &=& B^{i(1)}_{\,\in}(\bx)
   \frac{a^3_{\in}}{a^3(t)}~.
\eea
The average energy density of the first order magnetic field is then given by
\be\label{intensity}
\lan\rho_B^{(1)}\ran = \frac{1}{8\pi}\lan B_{{\rm (in)}}^{(1)2}({\bf x})\ran
                                   \frac{a^4_{\rm in}}{a^4(t)}   \;.
\ee
Here, we assume that the first order magnetic field has been generated
by some random process. Hence $B^{(1)}_{i\,\in}$ is a random variable
and $\lan\cdots\ran$ denotes the expectation
value. We assume also that this random process is statistically
homogeneous so that $\lan\rho_B^{(1)}\ran$ is independent of position.

\subsection{Gravitational waves}
For the gravity wave equation we consider a Fourier component
\bean
\si^{(1)k}_{ij}(\bx,t) &=& \si^{(1)}(\bk,t)Q_{ij}(\hat\bk)
   \exp(i\bk\cd\bx)  ~, \\
   D^2\si^{(1)k}_{ij}(\bx,t) &=& -\frac{k^2}{a^2(t)}\si^{(1)k}_{ij}(\bx,t)~.
\eean
Here $Q_{ij}(\hat\bk)$ is a transverse traceless polarization tensor.
 We assume that the gravity waves are statistically
isotropic and parity invariant so that both polarizations have the
same averaged square amplitudes.
For the amplitude $\si^{(1)}(\bk,t)$ we
obtain to first order the usual tensor perturbation propagation
equation (neglecting anisotropic stresses of the cosmic fluid)
\be
\ddot\si^{(1)}  +\left[k^2
  -\frac{3}{2}(1+w)\HH^2\right]\si^{(1)} =0 ~,
\ee
where $\HH=\dot a/a$ denotes the co-moving Hubble parameter,
$\HH=aH$, where $H$ is the physical Hubble parameter.
We now rewrite this equation in terms of the dimensionless variable
$\Sigma_{(1)}({\bf k},t) \equiv \sigma_{(1)}({\bf k},t)/(a_\in^2\Theta) =
\sigma_{(1)}({\bf k},t)/(3Ha_\in^2)$. We have normalized by the factor
$1/a_\in^2$ in order for the quantity $\Si$ to be independent of the
normalization of the scale factor. This is not true for $\si$ which
is $\si \propto a_\in^2$. In this way, $\Si$ can be directly related to
observable quantities which are of course independent of the normalization
of the scale factor. Equivalently, we will make use of the variable $\BB$ that is defined 
as $\BB\equiv \sqrt{\kappa}B/(3Ha_\in)$ in order to be independent of
the normalization of the scale factor, as well as $\Si$. 
In terms of $\Si$ the above equation becomes
\bea
&& \hspace{-3mm} \ddot\Sigma_{(1)}-3(1+w)\HH \dot\Sigma_{(1)}
 \nonumber \\   && \quad        +\left[k^2+\left( \frac{3}{2}
        +6w+ \frac{9}{2}w^2\right)\HH^2 \right]
        \Sigma_{(1)}
  =0~. \label{Si1}
\eea
In the matter or radiation era, the solutions to this linear
homogeneous differential equation are well known in terms of Bessel
functions. We are mainly interested in the radiation epoch, where
$w=1/3$. During radiation domination the Universe expands like
$a(t)\propto t$ such that $\HH =Ha =1/t$. We can therefore express the
scale factor as
\be\label{scal}
a(t) = H_{\in}a_{\in}^2t~.
\ee
In the radiation dominated Universe Eq.~(\ref{Si1}) reduces to
\begin{equation}
  \ddot\Sigma_{(1)}-\frac{4}{t}\dot\Sigma_{(1)}
  +\left(k^2+\frac{4}{t^2}\right)\Sigma_{(1)}=0   \;,
\end{equation}
with solution
\begin{equation}
  \Sigma_{(1)} \propto (kt)^3\left[j_1(kt)+
    y_1(kt)\right]~,
\end{equation}
where $j_n$ and $y_n$ denote the spherical Bessel functions of index
$n$~\cite{Abr}.

We distinguish the super- and sub-horizon behaviors. In
the long wavelengths limit, {$z\equiv kt \ll 1$},
we have
\begin{eqnarray*}
  \lim_{z\rightarrow 0} z^3 j_1(z) \simeq \frac{z^4}{3}  \;,
    \qquad
    \lim_{z\rightarrow 0} z^3 y_1(z) \propto -z  \;.
  \end{eqnarray*}
Taking into account only the faster growing mode, we obtain
\begin{equation}
    \label{Sigma_0 outside}
    \Sigma_{(1)}(t) \simeq \Sigma_{(1)}^{\rm in}
       \left(\frac{kt}{kt_{\rm in}}\right)^4   \;,
    \qquad kt\ll 1 \;,
  \end{equation}
  or equivalently
  \begin{equation}
    \Sigma_{(1)}(t) \simeq \Sigma_{(1)}^{\rm in}
     \left(\frac{a}{a_{\rm in}}\right)^4     \;,
    \qquad kt\ll 1 \;.
  \end{equation}
The quantity directly related to gravity waves is however given by
$\sigma_{(1)}=3Ha^2_\in\Sigma_{(1)}$, for which we obtain on
super-horizon scales
\begin{equation}  \label{Sigma_{(1)} outside}
    \sigma_{(1)}(t)\simeq \sigma_{(1)}^{\rm in} \left(\frac{a}{a_{\in}}\right)^2~,
    \qquad kt\ll 1 \;.
\end{equation}
  A direct consequence of this is that the
  ``gravity wave energy density'' is constant in time
  outside the horizon, as we show in the next sub-section. Of
  course the notion of ``gravity wave energy density'' and ``gravity
  wave'' is not strictly well defined for
  wavelengths larger than the size of the Hubble horizon. We shall just
  use the expression which is valid inside the horizon and call this
  the ``gravity wave energy density'' by analogy. It has a physical
  interpretation as a true energy density only once it enters the horizon.
  However, whenever this quantity becomes of the order of the
  background energy density, we know that perturbations become large
  and we can no longer trust linear perturbation theory.

Let us also consider the short wavelengths limit where $kt \gg 1$. In
this limit we can approximate
\begin{equation}
    \label{Sigma_{(1)} inside}
    \Sigma_{(1)}(t) \simeq  (kt)^2 \frac{\cos(kt)}{\cos(1)}
        \Sigma_{(1)}(kt=1) \;,
    \qquad kt\gg 1   \;,
  \end{equation}
  where the initial constant $\Sigma_{(1)}(kt=1)$ stands for the value
  of $\Sigma_{(1)}$ when it enters the horizon and can be obtained
  from Eq.~(\ref{Sigma_0 outside}),
  \[
     \Sigma_{(1)}(kt=1)\simeq \Sigma_{(1)}^{\rm in}
                        \left(\frac{1}{kt_{\rm in}}\right)^4  \;.
  \]
  The behavior of gravity waves
  on sub-horizon scales, $kt \gg 1$, is then given by
  \begin{equation}
  \sigma_{(1)}(t)\simeq 3a_\in^2H (kt)^2 \frac{\cos(kt)}{\cos(1)}
                       \Sigma_{(1)}(kt=1)   \;.
  \end{equation}
  We shall see that in this case the gravity waves energy density decreases
  like $1/a^4$, as it has to be for true gravity waves which are
  massless modes.

\subsection{Energy Densities \label{s:O1}}

As a first physically important quantity,
let us discuss  the energy densities of these first order perturbations
and the corresponding density parameters.

 The magnetic energy density is
 \begin{equation}
   \rho_B^{(1)} \equiv \frac{B_{(1)}^2}{8 \pi}=
      \frac{B_i^{(1)}B^{i}_{(1)}}{8 \pi}   \;,
 \end{equation}
with Eq.~(\ref{intensity}), this becomes
 \begin{equation}
   \rho_B^{(1)}(t)= \frac{1}{8 \pi} B_{(1)\,{\rm in}}^2
   \left(\frac{a^4_{\rm in}}{a^4}\right) \,.
 \end{equation}
In the radiation dominated universe under consideration, the density
parameter of the first order magnetic field is therefore given by
 \begin{equation}
   \Omega_B^{(1)}(t) \equiv \frac{\rho_B^{(1)}}{\rho_c}
       = \frac{8\pi G \rho_B^{(1)}}{3 H^2}
       = \frac{G}{3}\frac{B^2_{(1)\,{\rm in}}}{ H^2_{\rm in}}
       = \Omega_{B\,\in}^{(1)}\,.
 \end{equation}
 The density parameter $\Omega_B^{(1)}$
 is constant in time. Both, the background radiation and the magnetic
 field which is frozen in, scale in the same way with the
 expansion of the Universe.
As long as the magnetic field density parameter $\Omega_B^{(1)}$
is much smaller than $1$, the magnetic field can be considered a small
perturbation.

This is the result for a constant magnetic field. We also want to
consider a stochastic magnetic field. In this case $\bB(x)$ is a random
variable and its spectrum is given by~\cite{CD}
\bea
a^2(t)\bB(\bx,t) &=& \frac{1}{(2\pi)^3}
  \int {\rm d}^3k\bB(\bk)e^{i\bx\cd\bk}\,, \\
\hspace{-3mm} \lan B_i(\bk)B_j^*(\bq)\ran &=& \!(2\pi)^3\de(\bk\!-\!\bq)
\PP_{ij}(\hat\bk) \PP_{B\,\in}^{(1)}(k)~.
\eea
Here the basic time evolution of the magnetic field $\propto a^{-2}$ has been
removed so that, to first order $\bB(\bk)$ is independent of time.
$\PP_{ij}(\hat\bk)=\de_{ij}-k^{-2}k_ik_j$ is the
projection tensor  onto the plane normal to $\bk$. The
tensorial form of the spectrum is dictated by statistical
isotropy which also requires that $\PP_{B\,\in}^{(1)}$ depends 
only on the absolute
value $k=|\bk|$, and by the fact that $\bB$ is divergence free.
The Dirac delta is a consequence of statistical
homogeneity\footnote{One could also add a term which is odd under parity
but we disregard this possibility in this work~\cite{CDK}.}.
In this case we obtain
\bean
\lan\rho_B^{(1)}\ran &=&  \frac{1}{(2\pi)^68\pi}\int
{\rm d}^3k \int {\rm d}^3q \lan\bB(\bk)\bB(\bq)\ran e^{i\bx\cd(\bk-\bq)}  \\
  &=&   \frac{1}{(2\pi)^3}\int\frac{{\rm d}k}{k}k^3\PP_{B\,\in}^{(1)}(k)  \\
 &=& \int\frac{{\rm d}k}{k}\frac{{\rm d}\rho_B^{(1)}(k)}{{\rm d}\log k}~.
\eean
For the magnetic field density parameter at scale $k$ this yields
\be\label{rhoB1.spec}
\frac{{\rm d} \Omega_B^{(1)}(k,t)}{{\rm d}\log k} =  \frac{8\pi G}{3(2\pi)^3}
   \frac{k^3\PP_{B\,\in}^{(1)}(k)}{H_{\in}^2} =
  \frac{{\rm d} \Omega_{B\,\in}^{(1)}(k)}{{\rm d}\log k}~.
\ee

Let us now consider gravity waves.
The gravity wave energy density in real space is given by
\begin{equation}\label{rGW1}
  \rho_{GW}^{(1)}\equiv \frac{\lan\dot{h}_{ij}\dot{h}^{ij}\ran}{8 \pi G}
    \frac{1}{a^2} ~,
\end{equation}
where the factor $1/a^2$ comes from the fact that the dot denotes
the derivative with respect to conformal time and the difference of
a factor 4 in the normalization as compared \eg to \cite{Wein}
comes from our definition of the perturbation variable
[$g_{ij}=a^2(\de_{ij}+2h_{ij})$].  In Eq.~(\ref{rGW1}) $h_{ij}$ is considered
as tensor field with respect to the spatial metric $\de_{ij}$ so that there
are no scale factors involved in raising or lowering indices,
$h_{ij}={h_i}^j=h^{ij}$. For simplicity we shall keep this convention is this
section for all spatial tensors.

To lowest order the shear is given by
$\sigma_{ij}^{(1)} = a\dot{h}_{ij}$. Furthermore, the fact that
$\sigma_{ij}^{(1)}$ is transverse and traceless together with
statistical isotropy determines entirely the tensor structure of the
power spectrum.
\bean
&&  \lan \sigma_{ij}^{(1)\,\in}(\bk)\sigma_{lm}^{(1)\,\in}(\bq) \ran = \\
&&  \qquad\qquad
    (2\pi)^3\de(\bk-\bq)\MM_{ijlm}(\hat\bk)\PP_{\si\,\in}^{(1)}(k)~,
\eean
where~\cite{CD}
\bea
    {\cal M}_{ijlm}(\hat\bk)&\equiv&\de_{il}\de_{jm}+\de_{im}\de_{jl}
    -\de_{ij}\de_{lm}+ k^{-2}(\de_{ij}k_lk_m + \nonumber \\
    &&      \de_{lm}k_ik_j -\de_{il}k_jk_m - \de_{im}k_lk_j -\de_{jl}k_ik_m
    \nonumber \\   \label{e:MM}
    &&      -\de_{jm}k_lk_i) + k^{-4}k_ik_jk_lk_m~.
    \eea
We have ${\MM^{ij}}_{ij}=4$,
which takes into account the two polarization degrees of
freedom.
Therefore, considering that also for the shear
we do not multiply by the scale factor while
raising or lowering indices, $\sigma_{ij}=\sigma^{ij}$, we can
write the gravity waves energy density in terms of $\sigma_{ij}$ as
\bea
  \rho_{GW}^{(1)} = \frac{\lan\sigma_{ij}\sigma^{ij}\ran}{8 \pi G}
  \frac{1}{a^4}~.
\eea
For the contribution to the energy density per logarithmic
frequency interval we then obtain
\bea
 \frac{{\rm d}\rho_{GW}^{(1)}(k,t)}{{\rm d}\log k} &=& \frac{2 }{(2\pi)^3 G}
    \left[k^3\PP_{\si}^{(1)}(k,t)\right]
    \frac{1}{a^4}
    \nonumber \\
    &=& \frac{18 }{(2\pi)^3 G}
    \left[k^3\PP_{\Sigma}^{(1)}(k,t)\right] H^2
    \left(\frac{a_\in}{a}\right)^4   \;,
    \nonumber \\
    \label{rho_GW}
\eea
where we have used the relation $\sigma_{ij}=3Ha_\in^2 \Sigma_{ij}$ or
equivalently $\PP_{\si}^{(1)}(k,t)=9H^2a_\in^4\PP_{\Sigma}^{(1)}(k,t)$.
Finally, we can write the gravity wave density parameter as
\bea
  \frac{{\rm d}\Omega_{\rm GW}^{(1)}(k,t)}{{\rm d}\log k} &\equiv&  
       \frac{1}{\rho_c}
       \frac{{\rm d}\rho_{GW}^{(1)}}{{\rm d}\log k}
       \nonumber \\ 
       &=& \frac{48\pi}{(2\pi)^3} \left[k^3\PP_{\Sigma}^{(1)}(k,t)\right]
        \left(\frac{a_\in}{a}\right)^4    .
\eea

We have now to distinguish between super- and sub-horizon modes.
Using our super-horizon result for 
$\Sigma_{(1)}=\sigma_{(1)}/(3Ha_\in^2)$ where
$kt \ll 1$
\[
    \Sigma_{ij}^{(1)}(\bk,t) \simeq \Sigma_{ij\,\in}^{(1)}(\bk)
             \left(\frac{t}{t_{\rm in}}
             \right)^4 ,
\]
we find
\begin{eqnarray}
  \frac{{\rm d}\rho_{GW}^{(1)}(k,t)}{{\rm d}\log k} &=& \frac{18}{(2\pi)^3 G}
    \left[k^3\PP_{\Si\,\in}^{(1)}(k)\right]\left(\frac{a}{a_{\rm in}}\right)^8
    H^2   \left(\frac{a_\in}{a}\right)^4   \nonumber \\
    &=&\frac{18}{(2\pi)^3 G} \left[k^3\PP_{\Si\,\in}^{(1)}(k)\right]H_\in^2
    =\frac{{\rm d}\rho_{GW\,\in}^{(1)}}{{\rm d}\log k}~.
    \nonumber \\ 
\end{eqnarray}
For the last equal sign we made use of Eq.~(\ref{scal}). Hence on
super-horizon scales
the ``gravity wave energy density'' is time independent.
Then, of course the gravity wave density parameter grows like $a^4$,
\begin{equation}
 \label{e:OmGW1large}
  \frac{{\rm d}\Omega_{\rm GW}^{(1)}(k,t)}{{\rm d}\log k }
       = \frac{{\rm d}\Omega_{GW}^{(1)\,\in}}{{\rm d}\log k}
       \left(\frac{a}{a_{\rm in}}\right)^4\,, \quad kt\ll 1   \;,
\end{equation}
where
\begin{equation}
  \frac{{\rm d}\Omega_{GW}^{(1)\,\in}(k)}{{\rm d}\log k} =
  \frac{48\pi}{(2\pi)^3} \left[k^3\PP_{\Si\,\in}^{(1)}(k)\right]~.
\end{equation}
Inside the horizon, $kt \gg 1$, we have to insert
the expression of $\Sigma_{(1)}$ given by Eq.~(\ref{Sigma_{(1)}
inside}) in Eq.~(\ref{rho_GW}), which yields
\begin{eqnarray}
  \frac{{\rm d}\rho_{GW}^{(1)}(k,t)}{{\rm d}\log k} 
     &\simeq& \frac{9}{(2\pi)^3 G}
     \left[k^3\PP_{\Si\,\in}^{(1)}(k)\right]\frac{H_{\in}^2}{(kt_\in )^4}
   \left(\frac{a_\in}{a}\right)^4
   \nonumber \\
     &\propto& \frac{1}{a^4}~.
\end{eqnarray}
For the density parameter we obtain in a radiation dominated background
\bea
  \frac{{\rm d}\Omega_{GW}^{(1)}(k)}{{\rm d}\log k} &\simeq&  
  \frac{24\pi}{(2\pi)^3}
  \left(\frac{1}{k t_{\rm in}}\right)^4
  \left[k^3\PP_{\Si\,\in}^{(1)}(k)\right]   \nonumber \\ \label{omegaGWinside}
   &\simeq& \frac{1}{2}\left(\frac{1}{k t_{\rm in}}\right)^4
         \frac{{\rm d}\Omega_{GW}^{(1)\,\in}}{{\rm d}\log k}  \,,
         \quad kt\gg 1 ~.
\eea
Inside the horizon, the gravity wave density parameter is constant in
time as is natural in a radiation dominated Universe. Note that
this agrees, up to the factor $1/2$ which comes from averaging
$\cos^2(kt)$, with Eq.(\ref{e:OmGW1large}) at horizon entry, where
$(a/a_\in)^4 =(kt_\in)^{-4}$. Large scale gravity waves from
inflation, are ``amplified'' for a long time before entering the horizon,
\ie they have
$kt_\in\ll 1$.
Only if $\left[{\rm d}\Omega_{GW}^{(1)}(k)/{\rm d}\log k \right]$ 
is small for all values of $k$, 
 perturbation theory is justified. Therefore it is not sufficient if
 $\left[ {\rm d}\Omega_{GW}^{(1)\,\in}/{\rm d}\log k \right]$ 
is small, but we actually need that
$(k t_{\rm in})^{-4} \left[{\rm d}\Omega_{GW}^{(1)\,\in}/{\rm d}\log k\right]$
be small. This is better understood if we write the energy density in terms of
the metric perturbation. In a radiation dominated Universe the "growing" 
(not decaying) mode solution for the metric perturbation is
$$ h_{ij}(k,t)= e_{ij}(\bk)h_{\in}j_0(kt)~, $$
where $e_{ij}(\bk)$ is transverse traceless and $j_0$ is the spherical 
Bessel function of order $0$. Using $j_0' = -j_1$ and Eq.~(\ref{rGW1}) yields
$$  \rho_{GW}^{(1)} = k^3\frac{k^2\langle |h_\in|^2\rangle 
                                           j_1^2(kt)}{8\pi G a^2} $$
With $\rho_c = 3H^2/(8\pi G) = 3/(8\pi G a^2t^2)$ and $\langle |h_\in|^2
\rangle \equiv \PP_h$, we find
\be \label{k3Ph} 
      \frac{{\rm d}\Om_{GW}}{{\rm d}\log k} = 
        3[(kt)^2j_1^2(kt)]k^3P_h \simeq  3(kt)^4 k^3\PP_h ~,
         \quad \mbox{if } kt\ll 1~. 
\ee
Hence if the metric perturbations are small for all values of $k$, \ie
$k^3 \PP_h \ll 1$ this implies 
$$ \frac{{\rm d}\Om_{GW}}{{\rm d}\log k} \ll (kt)^4  ~.$$
Therefore the requirement $\left(k t_{\rm in}\right)^{-4}
\left[{\rm d}\Omega_{GW}^{(1)\,\in}/{\rm d}\log k\right] \ll 1$ is equivalent to the requirement 
that the metric perturbations be small on super horizon scales [note that 
$j_0(z) \simeq 1$ for $z\ll 1$].

Before we go to the second order, let us stress this point once more, 
because it is the origin of the confusion in the literature. Inflation 
generates gravitational waves with an amplitude
$$ k^3 \PP_h \simeq \left(\frac{H_{\mr{inf}}}{M_{\rm P}}\right)^2 \leq 10^{-10}~,$$
where $M_{\rm P}$ is the Planck mass and $H_{\mr{inf}}$ denotes the scale 
factor during inflation.  The maximum value of $10^{-10}$ is 
the maximum tensor fluctuation from inflation 
allowed by the cosmic microwave background anisotropies.

However, the density parameter on super-horizon scale is given by, 
see Eq.~(\ref{k3Ph})
$$ 
\frac{{\rm d}\Omega_{GW}^{(1)}}{{\rm d}\log k} \simeq (kt)^4 
\left(\frac{H_{\mr{inf}}}{M_{\rm P}}\right)^2,  \qquad kt\ll 1~.
$$
This equation is correct for any power law background, $a\propto t^q$, 
also for matter and even for inflation.
0nly at horizon crossing, can the density parameter become of the order 
$10^{-10}$. Inside the horizon it stays constant if  the background is 
radiation. Hence Eq.~(\ref{omegaGWinside}) can be written as
\be
  \frac{{\rm d}\Omega_{GW}^{(1)}(k)}{{\rm d}\log k} 
 \simeq         \left(\frac{H_{\mr{inf}}}{M_{\rm P}}\right)^2~,
         \qquad kt\gg 1 ~.
\ee


\section{Second order perturbations}
In this section we include all terms of second order in the perturbations, and
we shall insert our first order results for them;
\ie in terms of the form $\sigma_{ij}B^{j}$ we insert
$\sigma_{ij}^{(1)}B^{j}_{(1)}$  or for $\Pi_{ij}$ we insert the first order
magnetic fields, $\Pi_{ij}^{(1)}= B_i^{(1)}B_j^{(1)}-(1/3)\tilde 
p_{ij}^{(0)}B^{(1)2}$
in Eqs.~(\ref{eq. shear}) and~(\ref{eq. B}).
We obtain the following
differential equations for the evolution of the second order perturbations
$B_{i}^{(2)}({\bf x}, t)$ and $\sigma_{ij}^{(2)}({\bf x}, t)$:
\begin{eqnarray}
&& \hspace{-4mm}
\nabla_u\nabla_u{B}_{i}^{(2)} -D^2 B_{i}^{(2)} +\frac{5}{3}\Theta
        \nabla_u{B}_{i}^{(2)}
         + \frac{1}{3}\Theta^2 (1-w) B_{i}^{(2)} =
   \nonumber \\ &&  \quad
        \sigma_{ij}^{(1)} \nabla_u{B}^{j}_{(1)}+2\Theta
        \sigma_{ij}^{(1)} B^{j}_{(1)} 
        +2 \nabla_u{\sigma}_{ij}^{(1)}  B^{j}_{(1)}
  \nonumber \\ && \quad     \label{eq.B_1}
        +(D^2)^{(1)} B_{i}^{(1)}+ {\rm curl}J_i      \;,\\
&&  \hspace{-4mm}
\nabla_u\nabla_u{\sigma}_{ij}^{(2)}-D^2\sigma_{ij}^{(2)}
        +\frac{5}{3}\Theta \nabla_u{\sigma}_{ij}^{(2)}
        +\frac{1}{6}\Theta^2 (1-3w) \sigma_{ij}^{(2)} =
        \nonumber \\ && \quad
        \kappa\nabla_u{\Pi}_{ij}^{(1)}
        +\frac{2}{3}\Theta \kappa \Pi_{ij}^{(1)} +\Theta
        {\sigma_{\langle i(1)}}^{n}
        \sigma_{j\rangle n}^{(1)}+ (D^2)^{(1)} \sigma_{ij}^{(1)}
        \nonumber \\ && \quad \label{eq.shear_1}
        +2{\sigma_{\langle i}^{n (1)}
        \nabla_u\sigma_{j\rangle n}^{(1)}-
        \nabla_u{\sigma_{\langle i(1)}}^{n}}
        \sigma_{j\rangle n}^{(1)}     \;.
     \end{eqnarray}

Taking into account that $\nabla_uB_{i(1)} = -(2/3)\Theta B_{i(1)}$ together
with $\nabla_u{\Pi}_{ij}^{(1)} = -(4/3)\Theta {\Pi}_{ij}^{(1)} $,
Eqs.~(\ref{eq.B_1}), (\ref{eq.shear_1}) can be simplified to
\begin{eqnarray}
  \label{eq.B_1 bis}
&& \hspace{-4mm}
\nabla_u\nabla_u{B}_{i}^{(2)} -D^2 B_{i}^{(2)} +\frac{5}{3}\Theta
        \nabla_u{B}_{i}^{(2)}+
        \frac{1}{3}\Theta^2 (1-w) B_{i}^{(2)} =
        \nonumber \\ && \quad
        \left[ \frac{4}{3}\Theta \sigma_{ij(1)}
        +2\nabla_u{\sigma}_{ij}^{(1)}\right]  B^{j}_{(1)}
        +(D^2)^{(1)}B_{i}^{(1)} \;,
\\ &&\hspace{-4mm}
\nabla_u\nabla_u{\sigma}_{ij}^{(2)}-D^2\sigma_{ij}^{(2)}
        +\frac{5}{3}\Theta \nabla_u{\sigma}_{ij}^{(2)}
        +\frac{1}{6}\Theta^2 (1-3w) \sigma_{ij}^{(2)} =
    \nonumber \\ &&\quad
        -\frac{2}{3}\Theta \kappa \Pi_{ij}^{(1)} +\Theta
        {\sigma_{\langle i(1)}}^{n}
        \sigma_{j\rangle n}^{(1)}+ (D^2)^{(1)} \sigma_{ij}^{(1)}
        \nonumber \\ && \label{eq.shear_1bis} \quad 
        +2{\sigma_{\langle i}^{n (1)}
        \nabla_u\sigma_{j\rangle n}^{(1)}-
        \nabla_u{\sigma_{\langle i(1)}}^{n}}
        \sigma_{j\rangle n}^{(1)}     \;.
\end{eqnarray}
We have also neglected the term curl$J_i$ in Eq.~(\ref{eq.B_1}). Since 
it is proportional to $k$ in the Fourier space, its
contribution is important only on sub-horizon scales, where we anyway neglect
the source part. Outside the horizon,  $kt \ll 1$, it is negligible.

\subsection{The second order magnetic field from gravity waves and a
  constant magnetic field}
For simplicity, and to gain intuition, we first consider a constant
first order magnetic field,
\bean
  B_i^{(1)}({\bf x},t) &=& B^{(1)}_{i\,\in}\frac{a_{\rm in}}{a} ~,  \\
  B_i^{(1)}({\bf k},t)&=& B^{(1)}_{i\,\in} \frac{a_{\rm in}}{a}
\delta^3({\bf k})~.
\eean
In this case, the convolution of $B_{(1)}$ and $\si_{(1)}$ into which
the products in ordinary space transform under Fourier transformation,
become normal products and the second order magnetic
field $B^{(2)}_{i}$ has the same wavelength as the first order gravity wave
which generates it.

Remembering that $\si_{ij} \propto a^{-4}\sqrt{\PP_\si^{(1)}}\equiv
a^{-4}\si^{(1)}$ one obtains
\bea
&&  \ddot B_{i}^{(2)}+2\HH\dot B_i^{(2)}+B_i^{(2)}\left[k^2+\frac{1}{2}
  \HH^2 (1-3w)\right]=
\nonumber \\   && \qquad\qquad \qquad
2\dot{\si}^{(1)}_{ij} B_{j\,\in}^{(1)}
\frac{a_{\rm in}}{a^2} ~.
\eea
In principle, one has to consider the corrections to the orthogonal
spatially projected covariant derivative  $(D^2)^{(1)}B_{i}^{(1)}$
due to the tensor perturbations
$h_{ij}$ in the metric tensor $g_{\mu\nu}$.
Computing these corrections, they turn out to be equal to zero, since the
magnetic field is transverse. This remains valid even if $\bB^{(1)}$ is not
constant.

Considering the expansion-normalized dimensionless variable
$\BB_i^{(2)} \equiv \sqrt{\kappa}B_i^{(2)}/(\Theta a_\in) $,
we obtain
\begin{eqnarray}
  \label{eq.B_2/H const}
  &&\ddot\BB_i^{(2)}-\HH(1+3w)\dot\BB_i^{(2)}
  \nonumber \\ && \qquad
  +\BB_i^{(2)}\left[{k^2}
    + \HH^2 \left(\frac{1}{2}+3w+\frac{9}{2}w^2\right)\right]= f_i \,,
  \nonumber \\ &&
f_j\equiv  2\sqrt{\kappa}\left[\dot\Sigma_{ij}^{(1)}-\frac{3}{2}\HH (1+w)
    \Sigma_{ij}^{(1)} \right]  B^{(1)}_{j\, \in}
  \left(\frac{a_{\rm in}}{a}\right)^2 ~.
    \nonumber \\  \label{emB2}
\end{eqnarray}

We investigate the behavior of the second order perturbation in
the radiation dominated phase.

Moreover, since
the source $f_i(\bk,t)$ and therefore also $\BB_i^{(2)}({\bf k},t)$
are random variables, we want to determine their spectra.
The first order gravity wave spectrum is
\bean
&& \hspace{-5mm}
\langle \Sigma_{ij}^{(1)\,\in}({\bf k})
      \Sigma^{*(1)\,\in}_{ln}({\bf q}) \rangle
     = (2\pi)^3\mathcal{M}_{ijln}(\hat \bk) \delta^3({\bf k}-{\bf q})
       \PP^{(1)}_{\Sigma\,\in}(k)   \,,
  \nonumber \\ &&
   \langle \Sigma_{ij}^{(1)\,\in}({\bf k})
   \Sigma^{*ij}_{(1)\,\in}({\bf q}) \rangle
     = 4(2\pi)^3 \delta^3({\bf k}-{\bf q})
       \PP^{(1)}_{\Sigma\,\in}(k)   \,,
\eean
where $\mathcal{M}_{ijlm}$ is the gravity waves polarization tensor
defined in Eq.~(\ref{e:MM}). It can also be expressed in terms of the
projection tensor $\PP_{ij}(\hat\bk)$,
$\mathcal{M}_{ijlm} \equiv \mathcal{P}_{il}\mathcal{P}_{jm} +
\mathcal{P}_{im}\mathcal{P}_{jl}
- \mathcal{P}_{ij} \mathcal{P}_{lm}$. Actually $(1/2){\MM_{ij}}^{lm}$ is the
projection tensor onto the two transverse traceless modes of a rank 2 
symmetric tensor.
The power spectrum of the second order magnetic field $\BB_{(2)}$ is of
the form
\bea
&& \hspace{-4mm}
  \langle \mathcal{B}_{i}^{(2)}({\bf k},t)
      \mathcal{B}^{*(2)}_{j}({\bf p},t) \rangle
      = (2\pi)^3\PP_{ij}(\hat\bk)
      \delta^3({\bf k}\!-\!{\bf p})
      \mathcal{P}^{(2)}_{\BB}(k,t) \;.
      \nonumber \\
\eea

We obtain the solution for $\mathcal{B}_i^{(2)}({\bf k},t)$
with the help of Green function method,
\begin{equation}
  \BB_{i}^{(2)}({\bf k},t) = \int_{t_{\in}}^{t}{\rm d}t' \mathcal{G}
            (t,t',{\bf k}) f_i({\bf k},t')\,.
\end{equation}
Here $\GG$ is the Green function of the second order linear differential
operator acting on $\BB_{i}^{(2)}$ which depends on the cosmological
background. It can be determined in terms of the homogeneous solutions
which in the radiation dominated era are simply
spherical Bessel functions and powers. More precisely, in terms of
$z=kt$, Eq.~(\ref{emB2}) in the radiation dominated case, $w=1/3$, becomes
\be\label{b''z}
 \BB_{i}^{(2)\prime\prime} -\frac{2}{z} \BB_{i}^{(2)\prime} +
 \left( 1+\frac{2}{z^2}\right)  \BB_{i}^{(2)}
=k^{-2}f_i(z,\bk)~,
\ee
where the prime denotes a derivative w.r.t. $z$.
Two homogenous solutions to this equation are $P_1(z)=z^2j_0(z)$ and
$P_2(z)=z^2y_0(z)$. Defining the Wronskian,
$W(z)=P_1'(z)P_2(z)-P_1(z)P'_2(z)=z^2$, a possible Green function is
\be
\GG(z,z',\bk)=\frac{P_1(z')P_2(z)-P_1(z)P_2(z')}{W(z')}~.
\ee
The solution obtained by integrating with this Green function satisfies the
initial condition $  \BB_{i}^{(2)}(z_{\in},{\bf k}) =
\BB_{i}^{(2)\prime}(z_{\in},{\bf k}) =0$. Any other solution can be obtained by
adding a homogeneous solution to this one. We discuss the physically
correct choice of initial conditions in more detail in the Appendix
\ref{solution}. For the
magnetic field, the initial conditions chosen with this Green function seem
adequate to us. We can now write the magnetic field spectrum as
\bea
&& \langle \mathcal{B}_{i}^{(2)}({\bf k},t)
     \mathcal{B}^{*(2)}_{j}({\bf p},t) \rangle =
     \int_{z_{\in}}^{z}{\rm d}z'\int_{z_{\in}}^{z} {\rm d}z''
     k^{-2}p^{-2} \times \nonumber \\ &&
\quad\mathcal{G}(z,z',{\bf k})  \mathcal{G}^*(z,z'',{\bf p})
     \langle f_i({\bf k},z') f_j^*({\bf p},z'') \rangle     \;.
\eea

We solve Eq.~(\ref{b''z}),
distinguishing the sub- and super-horizon regimes.
In the long wavelength limit, $kt=z \ll 1$,
we have to insert the solution obtained
for gravity waves $\Sigma_{ij}^{(1)}$ on super-horizon scales and given in
Eq.~(\ref{Sigma_0 outside}).
Therefore, the source term $f_i(\bk, t)$ reads
\bea
  f_i(\bk,t')=4\sqrt{\kappa}\PP^s_i(\hat\bk)\left[\Sigma_{sn}^{(1)\,\in}(\bk)
    B_n^{(1)\,\in}\right] (H_{\in} a_{\in})^2 t'   \;,
\eea
and equivalently for $f^*_j(\bq,t'')$. The power spectrum of
$f_i$ can then be written as
\begin{eqnarray}
 &&  \langle f_i({\bf k},z') f_j^*({\bf p},z'') \rangle =
    \nonumber \\ && \qquad
    16\kappa \PP^s_i(\hat\bk)\PP^l_j(\hat\bq)
    \lan \Sigma_{sn}^{(1)\,\in}(\bk)\Sigma_{lr}^{*(1)\,\in}(\bk)\ran
    \nonumber \\ && \qquad
    B_n^{(1)\,\in}B_r^{*(1)\,\in} (H_{\in} a_{\in})^4 z' z''k^{-2}
    \nonumber \\ && \quad
    \equiv (2\pi)^3 \delta^3({\bf k-p})\mathcal{P}_{ij}(\hat\bk) h(z',z'',k)
    \;.
\end{eqnarray}
For the function $h(z',z'',k)$ we obtain the following expression
\bean
&&   h(z',z'',k)\simeq F(k) g(z') g(z'') \;, \\
&&   F(k)=\kappa B_{(1)\,\in}^2 \PP_{\Sigma\,\in}^{(1)}(k)k^{-2}  \;,\\
&&   g(z')=4 H_{\in}^2a_{\in} z'   \;.
\eean
The solution for the power spectrum of the second order perturbation
of the magnetic field can then be written as
\begin{eqnarray}
&&  \langle \mathcal{B}_{i}^{(2)}({\bf k},t) \mathcal{B}^{*(2)}_{j}({\bf p},t)
      \rangle
      = (2\pi)^3 \mathcal{P}_{ij}(\hat\bk) \delta^3({\bf k}\!-\!{\bf p})\times
\nonumber \\ &&
    \qquad  \left[ \int_{z_{\in}}^{z}{\rm d}z'
      \mathcal{G}(z,z',{\bf k})\sqrt{F(k)}g(z')\right]^2    \;.
\end{eqnarray}
The square $[\cdots]^2$ is simple the power spectrum
$\mathcal{P}_{\mathcal{B}}^{(2)}(k,t)$ which we want to determine. Of course,
the integrals in the square bracket are solutions to our magnetic field
Eq.~(\ref{b''z}) with source $\sqrt{F(k)}g(z)$. Hence
$\sqrt{\mathcal{P}^{(2)}_{\BB}}$ satisfies the equation
\bea
&&  P''- \frac{2}{z}P'+\left(
     1+\frac{2}{z^2}\right)P=\frac{\alpha}{k^3}z  ~, 
     \label{emP2}\\
&&  |P| \equiv \sqrt{\mathcal{P}_{\mathcal{B}}^{(2)}(k,t)} ~,  \qquad
    z\equiv kt ~,  \nonumber \\
&&  \alpha \equiv  4 H^2_{\in}a_{\in} \sqrt{\kappa B_{(1)\,\in}^2
    \PP_{\Sigma\,\in}^{(1)}(k)}  ~.  
    \nonumber
\eea
Solving the above equation with the Wronskian method
in the regime {$z=kt \ll 1$}, one finds
\bean
  P(z)\simeq \frac{\alpha}{2k^3}z^3 ~,
  \qquad z=kt \ll 1   ~.
\eean
This yields
\begin{eqnarray}
   k^3 \mathcal{P}_{\mathcal{B}}^{(2)}(k,t) &\simeq &
         4 \kappa \frac{B_{(1)}^{\in\,2}}{ H_{\rm in}^2}
         \left[k^3 \PP_{\Sigma\,\in}^{(1)}(k)\right] 
         \left(\frac{a}{a_{\rm in}}\right)^6 ~, 
         \nonumber \\ && \label{B_2/H outside}
         kt\ll 1~.
\end{eqnarray}
This is the second order magnetic field power spectrum
induced by the presence of a
first order field and a gravitational wave. It is the growth $\propto
t^6$ of this induced field which has been interpreted in
Refs.~\cite{Tsagas:2001ak,Tsagas:2005ki,D1} as strong amplification.
But before drawing such conclusions, we want to compare the energy density
parameter of $B^{(2)}$ with the one of $\si^{(1)}$ and $B^{(1)}$ inside
the horizon, where these quantities have a simple physical interpretation.

Inside the horizon,  {$kt \gg 1$},
we can no longer use the above simple
approximation for the source term. The solution of
 Eq.~(\ref{emP2}) with a generic source term,
\begin{eqnarray}
&&  \hspace*{-1cm} \left[\sqrt{\mathcal{P}_{\mathcal{B}}^{(2)}(k,z)}\right]''
        -\frac{2}{z}\left[\sqrt{\mathcal{P}_{\mathcal{B}}^{(2)}(k,z)}\right]'
        \nonumber \\&&  \qquad  \qquad
        + \left[\sqrt{\mathcal{P}_{\mathcal{B}}^{(2)}(k,z)}\right]=
        \mathcal{S}(k,z)   \;,
 \end{eqnarray}
can be written as
\be
 \sqrt{\mathcal{P}_{\mathcal{B}}^{(2)}(k,z)}=
      \int_{z_{\in}}^{z}dz'\mathcal{S}(k,z')\GG(z,z',\bk)\,.
  \ee
  But, once the gravity waves enter the horizon, the source and the 
Green function
  start oscillating and the contribution to the above integral
  becomes negligible. We therefore neglect the source inside the horizon
  and simply match the solution at horizon crossing with the homogeneous
  solutions of  Eq.~(\ref{emP2}) given above, that are 
  $P_1(z)=z^2j_0(z)$ and $P_2(z)=z^2y_0(z)$ ($z=kt$).
Considering the limit $z \gg 1$, this yields 
\begin{eqnarray}
  && k^3 \mathcal{P}_{\mathcal{B}}^{(2)}(k,t) \simeq 2\kappa
       \frac{B_{(1)}^{\in\, 2}}{H^2_{\rm in}}
         \left[k^3 \PP_{\Sigma\,\in}^{(1)}(k) \right]\times
         \nonumber \\ \label{B_1/H inside} && \qquad \quad
         \left(\frac{a}{a_{\rm in}}\right)^2
         \frac{1}{(kt_\in)^4}~,
        \quad   kt \gg 1 ~.
\end{eqnarray}

\subsubsection{The energy density}
To analyze this amplification which happens mainly on super-horizon
scales, let us compare energy densities after horizon entry.
The energy density of our second order magnetic field is
 \begin{eqnarray}
    \frac{{\rm d}\rho_{B}^{(2)}(k,t)}{{\rm d}\log k}
        &\equiv& \frac{1}{(2\pi)^3} \left[k^3 \PP_{B}^{(2)}(k,t)\right]
      \frac{1}{a^2}
        \nonumber \\
  \label{energy density B_2}
        &=& \frac{1}{(2\pi)^3} \left[k^3 \PP_{\BB}^{(2)}(k,t)\right]
        \frac{9H^2}{\kappa} \left(\frac{a_\in}{a}\right)^2 .
  \end{eqnarray}
The factor $1/a^2$ comes from the fact that we have to raise one index
of $\langle B_i^{(2)}B_i^{(2)}\rangle$ in order to compute the energy density, while $a_\in^2$ is 
due to the definition of $\BB_i^{(2)} \propto B_i^{(2)} /a_\in$ that we gave
above.  The density parameter for $B^{(2)}$ then reads
\begin{equation}
  \frac{{\rm d}\Omega_{B}^{(2)}(k,t)}{{\rm d}\log k}=\frac{3}{(2\pi)^3}
          \left(\frac{a_\in}{a}\right)^2
          \left[k^3 \mathcal{P}_{\mathcal{B}}^{(2)}(k,t) \right]  ~.
\end{equation}
With $H=H_{\in}a_{\in}^2/a^2$ we find that even though $\PP_{\BB}^{(2)}(k,t)$
is growing like $t^6$ on super-horizon
scales, the density
parameter grows like $ \Omega_{\rm GW}^{(1)}$. After
horizon entry, this growth stops and  $\Om_{B}^{(2)}$ remains constant.
Inserting the solutions (\ref{B_2/H outside}) and  (\ref{B_1/H inside}) for 
$k^3 \mathcal{P}_{\mathcal{B}}^{(2)}(k,t)$ gives
\begin{equation}\label{OmB2const}
  \frac{{\rm d}\Omega_{B}^{(2)}(k,t)}{{\rm d}\log k}= 6
          \frac{d\Omega_{\rm GW}^{(1)}(k,t)}{d\log k}
          \Omega_{B}^{(1)}
\end{equation}
on super- and sub-horizon scales.

Hence, even though the second order magnetic field $\BB_{(2)}$ is
growing considerably, this reflects only the growth of the unphysical 
density parameter $\Omega_{\rm GW}^{(1)}$ on super-horizon scales. Once 
this is factored in, the magnetic  field  density parameter is not. 
The values for both,
$\left[{\rm d}\Omega_{\rm GW}^{(1)\,\in}(k)/{\rm d}\log k\right](kt_\in)^{-4}  
=\left[{\rm d}\Omega_{\rm GW}^{(1)}(k)/{\rm d}\log k\right] $ and 
$ \Omega_{B}^{(1)}$ are at most of 
the order of $10^{-5}$ and smaller. For the gravity waves, we have seen 
that $\left[{\rm d}\Omega_{\rm GW}^{(1)\,\in}(k)/{\rm d}\log k\right]
(kt_\in)^{-4}$ is just 
the square amplitude of the
metric perturbations on super horizon scales, which has to be 
$k^3P_h\lsim 10^{-10}$ in order not to
overproduce Cosmic Microwavwe Background (CMB) anisotropies 
on large scales (integrated Sachs--Wolfe
effect). Similar arguments yield 
$\Omega_{B}^{(1)} < 10^{-5}$ on large scales (see, \eg~\cite{ped,DFK}). 
Therefore, even
 though we agree with the calculation in Ref.~\cite{Tsagas:2005ki}, we
 do not agree with the interpretation. If the gravitational wave energy
 density is as small as required by the measurements of CMB anisotropies,  
$\Omega_{B}^{(2)}$ always remains smaller than  $\Omega_{B}^{(1)}$.
 Furthermore, up to logarithmic corrections, $B^{(2)}$ inherits the spectrum
 of the first order gravity waves.

In the next section we
 show that this conclusion persists also if we allow for a stochastic
 magnetic field. Just the computation becomes more involved.

\subsection{The second order magnetic field from gravity waves and a
 stochastic magnetic field}

In the case in which the first order magnetic field is not spatially constant,
all the products $\Sigma_{ij}^{(1)}({\bf x},t) B^j_{(1)}({\bf x},t)$
become convolutions in Fourier space
\begin{eqnarray*}
&& \int {\rm d}^3x e^{i\bk\cd\bx} \Sigma_{ij}^{(1)}({\bf x},t) B^j_{(1)}({\bf x},t)=
\nonumber \\
 &&  \qquad      \frac{1}{(2\pi)^3}
      {\mathcal{P}_i}^n(\hat\bk)
      \int {\rm d}^3 q \Sigma_{nj}^{(1)}({\bf q},t)
      B^j_{(1)}({\bf k-q},t)      \;,
\end{eqnarray*}
where the projector ${\mathcal{P}_i}^n \equiv \delta_i^n-\hat{k}_i \hat{k}^n$
projects onto the transverse modes. The result of this
convolution is a magnetic field and therefore transverse. Hence this
projector is not strictly necessary. But as we shall see, it simplifies
the calculations.

Our equations are written in terms of
the dimensionless expansion-normalized variables
$\mathcal{B}_i^{(2)}({\bf x},t)$ and $\Sigma_{ij}^{(2)}({\bf x},t)$,
and we want to express their power spectra in terms of
the power spectra of the first order random variables
$B_i^{(1)}({\bf x},t)$ and $\Sigma_{ij}^{(1)}({\bf x},t)$ for which we assume
simple power laws,
\begin{eqnarray}
&& \hspace{-4mm}
   B_{i}^{(1)}({\bf k},t)= B_{i(1)}^{\in}({\bf k}) \frac{a_{\in}}{a}  \;,
   \nonumber \\ && \hspace{-4mm}
   B^i_{(1)}({\bf k},t)= B_{(1)}^{\in\, i}({\bf k}) \frac{a^3_{\in}}{a^3} \;,
   \nonumber \\ &&\hspace{-4mm}
   a_\in^2\langle B_{i}^{(1)\,\in}({\bf k}) B^{*(1)\,\in}_{j}({\bf q}) \rangle
        = (2\pi)^3 \mathcal{P}_{ij}(\hat\bk) \delta^3({\bf k}-{\bf q})
          \mathcal{P}^{(1)}_{B\,\in}(k)   \;,
   \nonumber \\ && \hspace{-4mm}
   \langle B_{i}^{(1)\,\in}({\bf k}) B^{*i}_{(1)\,\in}({\bf q}) \rangle
        = 2(2\pi)^3 \delta^3({\bf k}-{\bf q})
          \mathcal{P}^{(1)}_{B\,\in}(k)    \;,
   \nonumber \\ &&\hspace{-4mm}
   \mathcal{P}^{(1)}_{B\,\in}(k)= \left\{ \begin{array}{ll}
       [B_{(1)\,{\rm in}}^2 \lambda^3]   (\lambda k)^M
       & \textrm{ for $k<k_d$,} \\
       0 & \textrm{ for $k>k_d$,}
    \end{array} \right. \label{e:B1spec}
\end{eqnarray}
where $k_d$ is the damping scale which
we assume to be always much smaller than the Hubble scale. The scale 
$\la$ is arbitrary, e.g., the scale at which we want to calculate the
magnetic field. With this normalization $B_{(1)}^{\in}$ is simply
the amplitude of the magnetic field at scale $\la$ at time $t_\in$. At any other
moment, the magnetic field at scale $\la$ is given by
$B_{(1)}^{\in}a^2_\in/a^2(t)$.

Equivalently we have for the gravity wave power spectrum
\begin{eqnarray}
&& \Sigma_{ij}^{(1)}({\bf k},t)= \Sigma_{ij}^{(1)\,\in}({\bf k}) T(k,t) \;,
\nonumber \\ &&
   \langle \Sigma_{ij}^{(1)\,\in}({\bf k})
      \Sigma^{*(1)\,\in}_{ln}({\bf q}) \rangle
     = \nonumber \\ && \qquad
 (2\pi)^3\mathcal{M}_{ijln}(\hat\bk) \delta^3({\bf k}-{\bf q})
       \PP^{(1)}_{\Sigma\,\in}(k)    \;,
\nonumber \\ &&
   \langle \Sigma_{ij}^{(1)\,\in}({\bf k})
   \Sigma^{*ij}_{(1)\,\in}({\bf q}) \rangle
     = 4(2\pi)^3 \delta^3({\bf k}-{\bf q})
       \PP^{(1)}_{\Sigma\,\in}(k)    \;,
\nonumber \\ &&
    \PP^{(1)}_{\Sigma\,\in}(k)= [\Sigma_{(1)\,\in}^2 \lambda^3] (\lambda k)^A
       \;.
\label{2pGW}
\end{eqnarray}
Here the transfer function $T(k,t)$ keeps track of the deterministic
time-dependence of the gravity waves.
In the previous section we have derived the well known behavior of the
gravity wave transfer function which oscillates on sub-horizon
scales, $kt\gg1$, and behaves like a power law on super-horizon scales.
For the radiation dominated case,
\bea
   T(k,t) &\simeq& \left( \frac{a}{a_{\in}}\right)^4~,
       \label{tranf.func.} \qquad
       kt\ll 1~.
\eea

Starting from Eq.~(\ref{eq.B_1 bis}), we can write the following evolution
equation for the second order perturbation
\begin{eqnarray}
&&  \ddot B_{i}^{(2)}({\bf x},t)+2\HH\dot B_{i}^{(2)}({\bf x},t)
       -a^2 D^2 B_{i}^{(2)}({\bf x},t) +   \nonumber \\ && \qquad
       \frac{1}{2}\HH^2 (1-3w)B_{i}^{(2)}({\bf x},t) =
    2a\dot\sigma_{ij}^{(1)}({\bf x},t) B^j_{(1)}({\bf x},t)    \;.
    \nonumber \\
\end{eqnarray}
 Replacing $B_i^{(2)}=3Ha_\in \mathcal{B}_i^{(2)}/ \sqrt{\kappa}$
and $\sigma_{ij}^{(1)}=3Ha_\in^2  \Sigma_{ij}^{(1)}$, we obtain
\begin{eqnarray}
  && \ddot\BB_{i}^{(2)}({\bf x},t)-
       (1+3w)\HH\dot\BB_{i}^{(2)}({\bf x},t)
       -a^2 D^2 \mathcal{B}_{i}^{(2)}({\bf x},t) +
       \nonumber \\ && \qquad
       \left(\frac{1}{2}+3w+\frac{9}{2}w^2\right)
       \HH^2  \mathcal{B}_{i}^{(2)}({\bf x},t) =
  \nonumber \\ && \hspace*{-6mm}  \label{eq. B_1/H}
       2\sqrt{\kappa}a_\in a B^j_{(1)}({\bf x},t)
       \left[\dot\Sigma_{ij}^{(1)}({\bf x},t)-
       \frac{3}{2}\HH(1+w) \Sigma_{ij}^{(1)}({\bf x},t)\right] \,.
       \nonumber \\
\end{eqnarray}
This is the same differential equation as for the constant magnetic field.
In Fourier space this equation  becomes
\begin{eqnarray}
&&  \ddot\BB_{i}^{(2)}({\bf k},t)-
    (1+3w)\HH \dot\BB_i^{(2)}({\bf k},t)
    +\mathcal{B}_{i}^{(2)}({\bf k},t) \times
        \nonumber \\ && \qquad
  \left[k^2+\left(\frac{1}{2}+3w+\frac{9}{2}w^2\right)
    \HH^2 \right] =
    f_i({\bf k},t)\,,
\end{eqnarray}
where the source $f_i({\bf k},t)$ is now given by a convolution
\begin{eqnarray}
 &&  \hspace*{-4mm}
 f_i({\bf k},t)\equiv
       \frac{2}{(2\pi)^3}\sqrt{\kappa}a_\in a {\mathcal{P}_i}^r(\hat\bk) \times
 \nonumber \\   && \quad
       \left[\int {\rm d}^3 q  \dot\Sigma_{rj}^{(1)}({\bf q},t)
       B^j_{(1)}({\bf k-q},t)  -\frac{3}{2}(1+w)\HH \times
       \right.   \nonumber \\ && \quad \left.
       \int {\rm d}^3 q  \Sigma_{rj}^{(1)}({\bf q},t)
       B^j_{(1)}({\bf k-q},t)\right] ~ .
\end{eqnarray}

In terms of the variable $z=kt$ we obtain again Eq.~(\ref{b''z}). As in the
previous section  we solve it with the Green function method.
Therefore, the power spectrum of $\mathcal{B}_{i}^{(2)}$ is given by
\begin{eqnarray*}
 &&\langle \mathcal{B}_{i}^{(2)}({\bf k},t)
     \mathcal{B}^{*(2)}_{j}({\bf p},t) \rangle
     = (2\pi)^3\delta^3({\bf k}\!-\!{\bf p})\times 
         \\ && \qquad \quad
     (\delta_{ij}\!-\!\hat{k}_i \hat{k}_j) \mathcal{P}_{\BB}^{(2)}(k,t)~,
\end{eqnarray*}
with
\begin{eqnarray*}
  \mathcal{P}_{\BB}^{(2)}(k,t) &=& \int_{z_{\in}}^{z}{\rm d}z'
            \int_{z_{\in}}^{z} {\rm d}z''
            \mathcal{G}(z,z',{\bf k})\times 
                    \\ &&
            \mathcal{G}^*(z,z'',{\bf p})
            \langle f_i({\bf k},z') f_j^*({\bf p},z'') \rangle ~,
\end{eqnarray*}
where $z=kt$. In the radiation dominated epoch ($w= 1/3$) the
source term reads
\begin{eqnarray}
  && f_i({\bf k},t') =
       \frac{2}{(2\pi)^3}\sqrt{\kappa}a_\in a(t')
       {\mathcal{P}_i}^r(\hat\bk)\times  \nonumber \\  && \qquad
       \left[\int {\rm d}^3 q  \dot\Sigma_{rm}^{(1)}({\bf q},t')
       B^m_{(1)}({\bf k-q},t')-
       \right.  \nonumber \\  && \qquad \left.
       2\HH(t')
       \int {\rm d}^3 q  \Sigma_{rm}^{(1)}({\bf q},t')
       B^m_{(1)}({\bf k-q},t')\right]
       \nonumber \\  && \quad
       = \frac{2}{(2\pi)^3}\sqrt{\kappa}
       \frac{a^2_{\in}}{a^2(t')}
       {\mathcal{P}_i}^r(\hat\bk)  \times  \nonumber \\  && \qquad
       \left[\int {\rm d}^3 q  \Sigma_{rm}^{(1)\,\in}({\bf q})
         \dot T( q,t')
       B_{m}^{(1)\,\in}({\bf k-q})-
       \right.  \nonumber \\  && \qquad \left.
       2\HH(t') \!\!
       \int \! {\rm d}^3 q  \Sigma_{rm}^{(1)\,\in}({\bf q})
       T( q,t') B_{m}^{(1)\,\in}({\bk\!-\!\bq})\right] \;,
       \nonumber \\
\end{eqnarray}
and equivalently for $ f_j^*({\bf p},t'')$. To determine the power
spectrum of $f_i$ we assume that the magnetic field $B_{(1)}$ and
gravity waves $\sigma_{(1)}$  are uncorrelated, so that
\begin{eqnarray}
 && \hspace*{-4mm}
 \langle f_i({\bf k},t') f_j^*({\bf p},t'') \rangle =   \nonumber \\ && \quad
      \frac{16\kappa}{(2\pi)^6}
    \HH(t')\HH(t'')\left[\frac{a(t')a(t'')}{a^2_\in}\right]^2
    {\mathcal{P}_i}^r(\hat\bk){\mathcal{P}_j}^n(\hat\bp) \times
    \nonumber \\ &&  \quad 
    \int {\rm d}^3q \int  {\rm d}^3s
    \langle \Sigma_{rm}^{(1)\,\in}({\bf q})
    \Sigma^{*(1)\,\in}_{nl}({\bf s})\rangle
    \times  \nonumber \\ && \quad
    \langle B_{m}^{(1)\,\in}({\bk-\bq}) B^{*(1)\,\in}_l({\bf p-s})\rangle
    \nonumber \\ && \quad
    \equiv (2\pi)^3 \delta^3({\bk-\bp})\mathcal{P}_{ij}(\hat\bk) h(t',t'',k)~.
\end{eqnarray}
The function $h(t',t'',k)$ is given by \cite{CD}
\begin{eqnarray}
&& h(t',t'',k)=
     \frac{8\kappa}{(2\pi)^3}
     \HH(t')\HH(t'')\left[\frac{a(t')a(t'')}{a^2_\in}\right]^2 I(k)   \,,
     \nonumber \\ &&
  I(k) \equiv \int {\rm d}^3q (1+\gamma^2)(1+\alpha^2)\PP_{\Sigma\,\in}^{(1)}(q)
     \mathcal{P}^{(1)}_{B\,\in}
     (|{\bk-\bq}|)    \;,
\nonumber \\
\end{eqnarray}
where $\alpha\equiv \hat{k}\cdot (\widehat{k-q})$ and
$\gamma\equiv \hat{k}\cdot \hat{q}$.
We neglect the angular dependence of $(1+\gamma^2)$ and $(1+\alpha^2)$
and simply set
\[
   (1+\gamma^2)(1+\alpha^2)\simeq 1    \;.
\]
We then have to solve the following integral, 
\begin{eqnarray*}
  I(k) &=& 4\pi \Sigma_{(1)\,\in}^2 B_{(1)\,\in}^2 \lambda^{A+M+6}
        \int_0^{1/{\rm max}(t',t'')} \! {\rm d}q q^{A+2} \times
        \\ &&
        \int_{-1}^1 {\rm d} \mu (k^2+q^2-2\mu kq)^{M/2} ~.
\end{eqnarray*}
Here we evaluate the integral only up to the scale $q$ which enters
the horizon at the later of the two times. All scales
$q< 1/{\rm max}(t',t'')$ are super-horizon from $t_{\in}$ to
${\rm max}(t',t'')$. A soon as $q$ enters the horizon, the gravity wave
transfer function begins to oscillate and the contribution to the integral
becomes negligible. The integral over $\mu$ can be evaluated; for $M\neq -2$
it yields
\begin{eqnarray*}
   I(k) &=& \frac{8\pi}{2+M} \Sigma_{(1)\,\in}^2 B_{(1)\,\in}^2 \lambda^{A+M+6}
          \times
         \\ &&
         \int_0^{1/{\rm max}(t',t'')}\!\!
         \frac{ {\rm d}q~q^{A+2}}{kq}\left(|k\!+\!q|^{M+2}\! -\!
                       |k\!-\!q|^{M+2}\right)\, .
\end{eqnarray*}
We shall not treat the case $M=-2$, where the angular integral introduces
a logarithmic dependence on $q$, separately. This corresponds to
approximating $\log(k/q)\sim 1$. We approximate these integrals by
their dominant contribution.
\begin{itemize}
\item If the spectra are sufficiently red such that $A+M+3 <0$, the
      result is dominated by the region $k<1/{\rm max}(t',t'')$ and we obtain
    \begin{eqnarray*}
       I(k) &\simeq& 16 \pi \Sigma_{(1)\,\in}^2 B_{(1)\,\in}^2 \lambda^3
            \times \\ &&
            (\lambda k)^{A+M+3}\left(\frac{1}{A+3}-
             \frac{1}{A+M+3}\right)\,.
    \end{eqnarray*}
\item On the other hand, if the spectra are blue such that $A+M+3 >0$,
      the integral is dominated by its value at the upper boundary,
    \begin{eqnarray*}
      I &\simeq& 16\pi \Sigma_{(1)\,\in}^2 B_{(1)\,\in}^2 \lambda^{3}
      \frac{1}{A+M+3}
      \times \\ &&
      \left[\frac{\lambda}{{\rm max}(t',t'')}\right]^{A+M+3}~.
    \end{eqnarray*}
\end{itemize}

If, as in the previous sub-section, we can write the function
$h(t',t'',k)$ in the
form
\begin{equation}
  h(t',t'',k) \simeq
       F(k)g(t')g(t'')    \;,
\end{equation}
we can proceed as  we did before to obtain the results (\ref{B_2/H outside})
and (\ref{B_1/H inside}). A source where the time dependence of the unequal
time correlator factorizes is called ``totally coherent''. In the totally
coherent case, the power spectrum is simply the square of the solution which
has as its source the square root of the power spectrum of the
source~\cite{DKM}. In most cases, the unequal time correlator is more
complicated than this, but the totally coherent approximation is often
quite reasonable~\cite{DKM}.
If the source is totally coherent, the square root of the power spectrum
$\PP_{\BB}^{(2)}$ simply satisfies the same evolution equation as
$\BB_{(2)}$ with source term $\sqrt{F}g$.

\begin{itemize}
\item
 If $A+M+3 <0$, we can write
    \begin{eqnarray*}
       && F(k)= {\frac{128\pi\kappa}{(2\pi)^3}}
           (k\lambda)^{A+M+3}\lambda^3\times
              \nonumber \\ && \qquad
              \left(\frac{1}{A+3}-
             \frac{1}{A+M+3}\right)  \;,\\
       &&  g(t')= \frac{B_{(1)\,\in}\Sigma_{(1)\,\in}}{a^2_{\in}}
                    \HH(t') a^2(t') \;.
    \end{eqnarray*}
\item
  For  $A+M+3 >0$, we set
    \begin{eqnarray*}
       && F(k)=   {\frac{128\pi\kappa}{(2\pi)^3}}
                \frac{1}{A+M+3}\lambda^{A+M+6} \;,\\
       &&  g(t')=
                    \frac{B_{(1)\,\in}\Sigma_{(1)\,\in}}{a^2_{\in}}
                    \HH(t') a^2(t')
                    \left(\frac{1}{t'}\right)^{(A+M+3)/2}  \;.
    \end{eqnarray*}
This corresponds to replacing
\bean
\hspace{+4mm}
 \left[\frac{1}{\max(t',t'')}\right]^{(A+M+3)} \quad \mbox{ by }\quad
  \left(\frac{1}{t't''}\right)^{(A+M+3)/2} 
\eean
which is of course not entirely correct and we expect this to over estimate
the true result somewhat. However, within the accuracy of our approximations
this is sufficient. To obtain a more accurate result we would have to expand
the function $h(k,t',t'')$ in eigenfunctions with respect to convolution in
time, as it is done in Ref.~\cite{DKM}.
\end{itemize}

Within this totally coherent approximation we can now solve the problem
like in the previous sub-section. In the case {$A+M+3 <0$} we find
on super-horizon scales, where the
source is active
\begin{eqnarray}
    k^3\mathcal{P}^{(2)}_{\mathcal{B}}(k,t) &\simeq&
          \frac{ 32 \pi\kappa}{ (2\pi)^3}
          \frac{\left[k^3 \PP^{(1)}_{B\,\in}(k)\right]}{H_\in^2}
          \left[k^3 \PP^{(1)}_{\Sigma\,\in}(k)\right] \times
          \nonumber \\ &&
          \left(\frac{a}{a_{\in}}\right)^6
          \,, \quad kt\ll1  \;.
\end{eqnarray}
On sub-horizon scales, performing the matching at horizon crossing,  we obtain
\begin{eqnarray}
    k^3\mathcal{P}^{(2)}_{\mathcal{B}}(k,t) &\simeq&
          \frac{16\pi\kappa}{(2\pi)^3}
          \frac{\left[k^3 \PP^{(1)}_{B\,\in}(k)\right]}{ H_\in^2}
          \left[k^3 \PP^{(1)}_{\Sigma\,\in}(k)\right] \times
          \nonumber \\ &&
          \label{P_B2 red inside}  
          \left(\frac{a}{a_{\in}}\right)^2
          \frac{1}{(kt_\in)^4}
          ~,  \quad kt\gg1 ~.
\end{eqnarray}

If {$A+M+3 >0$}, we analyze in in more
detail only the case {$A\simeq -3$} and $M=2$. The spectral index $A=-3$
correspond to a scale invariant  gravity wave power spectrum as it is
obtained in slow-roll inflation~\cite{Book}. The index
$M=2$ characterizes a causal magnetic field $B_{(1)}$.
In this case, we have to solve the differential equation,
\bea
     \label{eq.B}
  && P''-\frac{2}{z}P'+\left(1+\frac{2}{z^2}\right)P=\frac{\alpha}{k^2}   ~, \\
  && \alpha \equiv
          (a_\in H_\in)^2 B_\in^{(1)} \Sigma_\in^{(1)}
          \sqrt{
          \frac{64\pi}{ (2\pi)^3}\kappa \lambda^5}
             ~,  \nonumber
\eea
where the source is constant in time. Detailed comments about the
initial conditions chosen for the solution
of the above equation can be found in
Appendix~\ref{solution}. Finally, we can write the solution for $P(z)$
in the case where $z=kt \ll 1$ as
\bean
  P(z)\simeq \frac{\alpha}{ k^2}z^2 \log\left(\frac{z}{z_\in}\right)
  ~, 
  \qquad z \ll 1~.
\eean
The power spectrum of $\mathcal{B}^{(2)}$ on super-horizon scales is
therefore given by
\begin{eqnarray}
  &&  k^3\mathcal{P}_{\mathcal{B}}^{(2)}(k,t) \simeq
          \frac{16\pi\kappa}{ (2\pi)^3}
          \frac{\left[k^3 \PP^{(1)}_{B\,\in}(k)\right]}{H_\in^2}
          \left[k^3 \PP^{(1)}_{\Sigma\,\in}(k)\right] \times
          \nonumber \\ && \qquad
          \left(\frac{a}{a_{\in}}\right)^4
          \frac{1}{(kt_\in)^2} \log^2\left(\frac{a}{a_\in}\right)
          \,, \qquad kt \ll 1~.
\end{eqnarray}
On sub-horizon scales, $z=kt \gg 1$, we match 
the super-horizon solution at horizon
crossing with the homogeneous solution of 
Eq.~(\ref{eq.B}), as we did above, obtaining
\begin{eqnarray}
  &&  k^3\mathcal{P}_{\mathcal{B}}^{(2)}(k,t) \simeq \frac{32\pi\kappa}{
          (2\pi)^3}
          \frac{\left[k^3 \PP^{(1)}_{B\,\in}(k)\right]}{H_\in^2}
          \left[k^3 \PP^{(1)}_{\Sigma\,\in}(k)\right] \times
          \nonumber \\ && \qquad 
          \left(\frac{a}{a_{\in}}\right)^2
          \frac{1}{(kt_\in)^4} \log^2\left(kt_\in\right)
          \,, \qquad kt \gg 1~.
\end{eqnarray}

\subsubsection{Density parameter}
Using Eq.~(\ref{energy density B_2}),
we find the following expressions for the energy density of the
stochastic second order magnetic field.
If $A+M+3 <0$, we have on super-horizon scales
\begin{eqnarray}
 \frac{{\rm d}\rho_{B}^{(2)}(k,t)}{{\rm d}\log k} &\equiv&
       \frac{1}{(2\pi)^3}k^3 \mathcal{P}^{(2)}_{B}(k,\eta)
       \left(\frac{a_\in}{a}\right)^2
       \nonumber \\
       &\simeq&  \frac{288\pi }{(2\pi)^6}
       \left[k^3 \PP^{(1)}_{B\,\in}(k)\right]
       \left[k^3 \PP^{(1)}_{\Sigma\,\in}(k)\right]    \,.
\end{eqnarray}
This results in a density parameter for $B_{(2)}$ given by
\begin{eqnarray}
   \frac{{\rm d}\Omega_{B}^{(2)}(k,t)}{{\rm d}\log k}
       &\equiv& \frac{1}{\rho_c} \frac{d\rho^{(2)}_{B}(k,\eta)}{d\log k}
       \nonumber \\
       &\simeq&  6
       \frac{{\rm d}\Omega^{(1)}_{B\,\in}(k)}{{\rm d}\log k}
       \frac{{\rm d}\Omega^{(1)}_{GW}(k,t)}{{\rm d}\log k}
       ~,
           \nonumber \\ && kt \ll 1     ~. \label{e:B2red}
\end{eqnarray}
Inside the horizon we obtain for the second order magnetic field 
density parameter
\begin{eqnarray}
   \frac{{\rm d}\Omega_{B}^{(2)}(k,t)}{{\rm d}\log k}
       &\simeq&  6
       \frac{{\rm d}\Omega^{(1)}_{B\,\in}(k)}{{\rm d}\log k}
       \frac{d\Omega^{(1)}_{GW}(k)}{d\log k}
       ~,  
           \nonumber \\ && kt \gg 1 ~.
\end{eqnarray}
The gravity wave density parameter, 
$\left[{\rm d}\Omega^{(1)}_{GW}(k,t)/{\rm d}\log k\right]$ is given
by Eqs.~(\ref{e:OmGW1large}) and~(\ref{omegaGWinside}) respectively.
This corresponds, as in the previous section for a constant magnetic
field, to the naively expected result, 
$\Om_{B}^{(2)} \sim \Om^{(1)}_{\rm GW}\Om^{(1)}_{B}$.

For blue spectra, $A+M+3>0$, the second order magnetic field density
parameter reads in the interesting case
$A\simeq -3$ and $M=2$ on super-horizon scales
\begin{eqnarray}
   \frac{{\rm d}\Omega^{(2)}_{B}(k,t)}{{\rm d}\log k}
       &=&  \frac{12}{(kt)^2}
       \frac{{\rm d}\Omega^{(1)}_{B\,\in}(k)}{{\rm d}\log k}
       \frac{{\rm d}\Omega^{(1)}_{\rm GW}(k,t)}{{\rm d}\log k} 
       \log^2\left(\frac{a}{a_\in}\right)~ 
       \nonumber \\
       &=& 12  \frac{{\rm d}\Omega^{(1)}_{\rm GW}(k,t)}{{\rm d}\log k} 
       \left.\frac{{\rm d}\Omega_{B\,\in}^{(1)}(k)}{{\rm d}\log k} 
       \right|_{k=1/t} (kt)^{3}
       \times
       \nonumber \\ &&  \label{e:B2blue}
       \log^2\left(\frac{a}{a_\in}\right)~, 
       \quad kt \ll 1 ~.
\end{eqnarray}
Note that the value of 
$ \left[{\rm d}\Omega^{(2)}_{B}(k,t)/{\rm d}\log k\right]$
on super-Hubble scales is affected by 
$ \left[{\rm d}\Omega^{(1)}_{B}(k_t)/{\rm d}\log k_t \right]$ 
at horizon crossing, 
$k_t=1/t$ which may well be larger than 
$\left[{\rm d}\Omega^{(1)}_{B}(k)/{\rm d}\log k \right]$ but of course has 
also to be much smaller than $1$. 

This expression grows only logarithmically faster than
$\left[{\rm d}\Omega_{GW}^{(1)}(k,t)/{\rm d}\log k\right]$. 
The growth stops at horizon entry where the
second order magnetic field density parameter has acquired a factor
$\log^2(kt_\in)$.
Inside the horizon we obtain a density parameter of
\begin{eqnarray}
   \frac{{\rm d}\Omega^{(2)}_{B}(k,t)}{{\rm d}\log k}
       &=& 12\frac{{\rm d}\Omega^{(1)}_{B\,\in}(k)}{{\rm d}\log k}
       \frac{{\rm d}\Omega^{(1)}_{\rm GW}(k)}{{\rm d}\log k}
       \log^2 (kt_\in)~,
           \nonumber \\ && kt \gg 1 ~.
\end{eqnarray}
Up to the logarithmic correction, this corresponds to the result for red
spectra above.

\subsubsection{Reheating and matter dominated epochs}

In order to make contact with Refs.~\cite{Tsagas:2001ak,Tsagas:2005ki}, we now
repeat the calculation in a matter dominated background ($w=0$).
We want to point out that the results we obtain are mathematically the same
as the ones found in \cite{Tsagas:2005ki}. The only difference lies in the
interpretation. In the previous paragraph we have seen that, even though
$$
 \frac{{\rm d}\Omega_{GW}^{(1)}(k)}{{\rm d}\log k} \sim  
    \left(\frac{1}{k t_{\rm in}}\right)^4
  \left[k^3\PP_{\Si\,\in}^{(1)}(k)\right]  ~,
$$
and even though $(kt_\in)^{-4}$ can become very large, this product is 
never larger than about $10^{-10}$. We believe that this point has been 
missed in Ref.~\cite{Tsagas:2005ki}.
\\
If $w=0$, the scale factor grows like $a\propto t^2$ so that
$\HH =2/t$. 
As mentioned before, for the super horizon amplification the question 
whether the conductivity is high or low is not relevant.

>From the first order perturbations, we obtain the same
behaviour for the magnetic field $B^{(1)}$ in terms of the scale factor,
therefore the density parameter is then given by
 \bea 
  \frac{{\rm d}\Omega^{(1)}_B(k,t)}{{\rm d}\log k} = \frac{8\pi G}{3(2\pi)^3}
     \frac{\left[k^3 \PP_{B\,\in}^{(1)}(k) \right]}{H_\in^2}
     \frac{a_\in}{a} ~.
\eea
The first order gravity waves on super-horizon scales now behaves as
\bea
   \Sigma_{ij}^{(1)}(\bk,t)=\Sigma_{ij\,\in}^{(1)}(\bk)
       \left( \frac{a}{a_\in} \right)^3    ~.
\eea
Once the gravitational waves enter the horizon, they start oscillating and the
energy density decays as radiation. Therefore in this case the relative 
density parameters for the first order gravity waves is
\bea 
  \frac{{\rm d}\Omega^{(1)}_{\rm GW}(k,t)}{{\rm d}\log k} &=& 
     \frac{48\pi}{(2\pi)^3}
     \left[k^3 \PP_{\Si\,\in}^{(1)}(k) \right] 
     \left( \frac{a}{a_\in} \right)^2  ~, 
     \nonumber  \\
     && kt \ll 1 ~.
\eea
 On sub-horizon scales we obtain
\bea 
  \frac{{\rm d}\Omega^{(1)}_{\rm GW}(k,t)}{{\rm d}\log k} &=& 
     \frac{24\pi}{(2\pi)^3}
     \left[k^3 \PP_{\Si\,\in}^{(1)}(k) \right] 
     \left( \frac{a_\in}{a} \right)  \frac{1}{(kt_\in)^6} ~, 
     \nonumber  \\
     && kt \gg 1 ~.
\eea

Computing finally the induced second order magnetic field density parameter,
we obtain the naively expected result on super-horizon scales
\begin{eqnarray}
   \frac{{\rm d}\Omega^{(2)}_{B}(k,t)}{{\rm d}\log k}
       &\simeq& \left\{ \begin{array}{ll} 
\frac{{\rm d}\Omega^{(1)}_{B}(k,t)}{{\rm d}\log k}
       \frac{{\rm d}\Omega^{(1)}_{\rm GW}(k,t)}{{\rm d}\log k}~, & \\
   \qquad \mbox{for } A+M+3<0 & \\
 (kt)^3\left[\frac{{\rm d}\Omega^{(1)}_{B}(k,t)}{{\rm d}\log k}\right]_{k=1/t}
     \times & \nonumber \\
      \qquad \frac{{\rm d}\Omega^{(1)}_{\rm GW}(k,t)}{{\rm d}\log k}
          \log^2\frac{a}{a_\in}~,
       &
  \nonumber \\
 \qquad \mbox{for } A+M+3> 0 & \\ \end{array} \right.\\ 
           \nonumber \\ && kt \ll 1 ~.
\end{eqnarray}
 On  sub-horizon scales
the density parameter turns out to be given by 
\begin{eqnarray}
   \frac{{\rm d}\Omega^{(2)}_{B}(k,t)}{{\rm d}\log k}
       &\simeq&
\frac{{\rm d}\Omega^{(1)}_{B}(k,t)}{{\rm d}\log k}
       \frac{{\rm d}\Omega^{(1)}_{\rm GW}(k,t_k)}{{\rm d}\log k}
           \nonumber \\ 
 &\simeq& \frac{{\rm d}\Omega^{(1)}_{B}(k,t)}{{\rm d}\log k} 
         \left(\frac{H_{\mr{inf}}}{M_{\rm P}}\right)^2 , ~ kt \gg 1\, , 
\end{eqnarray}
for both cases $A+M+3<0$ and $A \simeq -3$, $M=2$, 
up to logarithmic corrections. Here $t_k$ stands for the horizon crossing
time, $t_k=1/k$, and in the last $\simeq$ sign we have used that
$\left[{\rm d}\Omega^{(1)}_{\rm GW}(k,t_k)/{\rm d}\log k\right] \simeq
(H_{\rm{inf}}/M_{\rm P})^2 $ is the
gravity waves density parameter at horizon crossing, which is 
smaller than $10^{-10}$. This means that the second order magnetic field does 
not grow larger the the first order one. Inside the
horizon they decrease both like 
$\propto a^{-1}$. $\Omega_B^{(2)}$ stays always much
smaller than $\Omega_B^{(1)}$, as we have found in the case of a radiation
dominated background.


\subsection{Second order gravity waves}

Starting from Eq.~(\ref{eq.shear_1bis}), we can write the evolution equation
for $\sigma_{ij}^{(2)}$ in real space $({\bf x},t)$ as follows:
\begin{eqnarray}
  &&  \ddot{\sigma}_{ij}^{(2)}-a^2 D^2\sigma_{ij}^{(2)}
        -\frac{3}{2}\HH^2(1+w)  \sigma_{ij}^{(2)} =
        \nonumber \\ && \qquad \qquad
        -2 \kappa a \HH\Pi_{ij}^{(1)} +\left[a \HH
        {{\sigma_{\langle i(1)}}^{n}}
        {\sigma^{(1)}_{j\rangle n}}+ \right.
        \nonumber \\ && \qquad \qquad \left.
        2a{{\sigma_{\langle i(1)}}^{n}}
        {\dot{\sigma}^{(1)}_{j\rangle n}}- a
        {{\dot{\sigma}_{\langle i(1)}}^{n}}
        {\sigma^{(1)}_{j\rangle n}}\right]
        \frac{1}{a^2}   \;.
        \nonumber \\
\end{eqnarray}
The factor $1/a^2$ in the source part of the above equation
comes from the fact that in Eq.~(\ref{eq.shear_1bis}) we had to add
factors $a^2(t)$ in order to lower or rise indices.
On the other hand, now we deal with purely
spatial tensors such that $\sigma_{ij}=\sigma^{ij}$ and also
$\dot\sigma_{ij}=\dot\sigma^{ij}$.

Introducing again the dimensionless expansion-normalized variable
$\Sigma_{ij}^{(2)}$, the previous equation can be written as
\begin{eqnarray}
 && \ddot{\Sigma}_{ij}^{(2)}-3(1+w)\HH\dot{\Sigma}_{ij}^{(2)}
        +3\HH^2 \left( \frac{3}{2}w^2
        +2w+ \frac{1}{2}\right)
        \Sigma_{ij}^{(2)} \nonumber \\  &&\quad
        -a^2D^2\Sigma_{ij}^{(2)}=
        -\frac{2}{3}\kappa \frac{a^2}{a_\in^2} \Pi_{ij}^{(1)} +
        \nonumber \\  &&\quad
        \left[-\frac{3}{2} (1+3w)\HH^2
        {{\Sigma_{\langle i(1)}}^{n}}
        {\Sigma^{(1)}_{j\rangle n}} \right.
        \nonumber \\ && \qquad \left.
        +6\HH{{\Sigma_{\langle i(1)}}^{n}}
        {\dot{\Sigma}^{(1)}_{j\rangle n}}- 3\HH
        {{\dot{\Sigma}_{\langle i(1)}}^{n}}
        {\Sigma^{(1)}_{j\rangle n}}\right]
            \left(\frac{a_\in}{a}\right)^2      \;. \label{emS2}
\end{eqnarray}
As for $B^{(2)}$, the source is given
by the first order perturbations magnetic field [$\Pi_{ij}^{(1)}$] and
the first order gravity waves and does \eg not couple to the second order
magnetic field. Since we assume the first order magnetic field and gravity wave
fluctuations to be independent, we can add the power spectra for the
solutions of the individual source terms,
\bean
  k^3\PP_\Sigma^{(2)}(k,t)= k^3\PP_\Sigma^{(2)\,\Pi}(k,t)+
                           k^3\PP_\Sigma^{(2)\,{\rm GW}}(k,t)   \;.
\eean
where $\PP_\Sigma^{(2)\,\Pi}(k,t)$ is the power spectrum of the solution
of Eq.~(\ref{emS2}) with source  term $\Pi^{(1)}$ only and
$\PP_\Sigma^{(2)\,{\rm GW}}(k,t)$ 
comes from the source terms containing $\Si^{(1)}$.

\subsubsection{Magnetic field part of the source   \label{s:O2Pi}
$\left[k^3\PP_\Sigma^{(2)\,\Pi}(k,t)\right]$  }

  Considering first the magnetic field part of the source, we have to
  solve the following differential equation in the momentum space
  $({\bf k},t)$ 
  \bea\label{emSi2}
   &&\ddot{\Sigma}_{ij}^{(2)}-3(1+w)\HH\dot{\Sigma}_{ij}^{(2)}
        +\Bigg[k^2+
        \nonumber \\  &&\quad
        3\HH^2 \left( \frac{3}{2}w^2
        +2w+ \frac{1}{2}\right)\Bigg]
        \Sigma_{ij}^{(2)} =
         f_{ij}   \;,
  \eea
  where the source is given by
  \bea
     f_{ij}({\bf k},t)\equiv -\frac{2}{3}\kappa \frac{a^2}{a_\in^2}
        \Pi_{ij }^{(1)}({\bf k},t) \;.
  \eea
  As before, we have to compute the unequal time correlator:
  \bea
  && \langle\Pi_{ij}^{(1)} ({\bf k},t')\Pi_{rn}^{*(1)}({\bf p},t'') \rangle
  =  \nonumber \\ && \qquad
  (2\pi)^3 \delta^3({\bk-\bp}) \MM_{ijrn}({\bf \hat k})
  h(k,t',t'')  \;,
  \label{expValPi}
  \eea
  where the anisotropic stresses are given by
  \bean
    \Pi_{ij}^{(1)} ({\bf k},t')=-\frac{1}{16\pi(2\pi)^3} {\MM_{ij}}^{ls}
      ({\bf \hat k})
      \int {\rm d}^3q B_l^{(1)}({\bf q},t')\times &&  \\
      B_s^{(1)}({\bk-\bq},t')   &&  \,.
  \eean
  $(1/2){\MM_{ij}}^{ls}({\bf \hat k})$ is the projector on the tensor modes.
  We have neglected a trace contribution to the magnetic field stress tensor
  since, once we project with ${\MM_{ij}}^{ls}$,  the trace  vanishes.

  After some computation~\cite{CD}, we find for the function $h(k,t',t'')$
  the following expression:
  \bea
    \label{h}
   h(k,t',t'') &=& \frac{1}{(8\pi)^2}\frac{1}{4(2\pi)^3}  I(k)
      \left[\frac{a_\in^2}{a(t')a(t'')}\right]^2    ,  \\
    I(k) &=& \int {\rm d}^3q (1+\gamma^2)(1+\alpha^2) \PP_{B\,\in}^{(1)}(q)\times
    \nonumber \\ && \qquad\qquad
      \PP_{B\,\in}^{(1)}(|{\bk-\bq}|)    \;.
  \eea
  where $\alpha\equiv \hat{k}\cdot (\widehat{k-q})$ and
  $\gamma\equiv \hat{k}\cdot \hat{q}$.
  As before, we approximate $(1+\gamma^2)(1+\alpha^2) \simeq 1$.
  With this, we obtain the following expression for the expectation
  value of the source term:
  \bea
  &&\langle f_{ij} ({\bf k},t') f_{rn}^*({\bf p},t'') \rangle
      =\frac{4}{9}\kappa^2 \frac{a^2(t')a^2(t'')}{a_\in^4} \times
      \nonumber \\ && \qquad \quad
      \langle\Pi_{ij}^{(1)} ({\bf k},t')\Pi_{rn}^{*(1)}({\bf p},t'')
      \rangle   \;.
      \label{expValSource}
  \eea
The expectation value of the stochastic
variable $\Sigma_{ij}^{(2)}$ can be written as
\bea
  && \langle\Sigma_{ij}^{(2)} ({\bf k},t)
  \Sigma_{rn}^{*(2)}({\bf p},t) \rangle
  =  \nonumber \\ && \qquad
  (2\pi)^3 \delta^3({\bf k-p}) \MM_{ijrn}({\bf \hat k})
  \PP_{\Sigma}^{(2)} (k,t) \;.
\eea
If $\langle\Pi_{ij}^{(1)} ({\bf k},t')\Pi_{rn}^{*(1)}({\bf p},t'')\rangle$
can be written as a product of a function of $(k,t)$ and $(k,t'')$, this 
source is totally coherent and we can write the function
$h(k,t',t'')$ of  Eq.~(\ref{h}) in the form
\bean
    \frac{4}{9}\kappa^2 \frac{a^2(t')a^2(t'')}{a_\in^4}h(k,t',t'')= 
F(k)g(t')g(t'') ~,
\eean
where we introduced the pre-factor of $h$ since we finally need an expression
for the unequal time correlator of the source, as in Eq.~(\ref{expValSource}),
while the function $h$ alone is only part of the correlator of the
anisotropic stress, Eq.~(\ref{expValPi}). \\
The square root of the power spectrum is then a solution of the
differential equation (\ref{emSi2}) with source term $\sqrt{F(k)}g(t)$. Written as
differential equation for the variable $z=kt$ and setting $w=1/3$, this becomes
  \bea
  && \hspace*{-0.6cm}\left[\sqrt{\mathcal{P}_{\Sigma}^{(2)\,\Pi}(k,z)}\right]''
      -\frac{4}{z}\left[\sqrt{\mathcal{P}_{\Sigma}^{(2)\,\Pi}
       (k,z)}\right]'
       \nonumber \\ &&  \hspace*{-0.4cm}
       + \left(1 +\frac{4}{z^2}\right)
       \left[\sqrt{\mathcal{P}_{\Sigma}^{(2)\,\Pi}(k,z)}\right]=
       \sqrt{F(k)}\frac{g(z/k)}{k^2}   \;.   \label{eq.Sigma_2}
  \eea

  As for the second order magnetic field, we distinguish between two cases.
First we consider $2M+3>0$. The integral $I$ is then dominated by the upper
cutoff. The magnetic field is not oscillating and we therefore take 
damping scale $k_d$ as the upper cutoff. We neglect the slow time dependence
of this scale. Using Eq.~(\ref{e:B1spec}) for the magnetic field power spectrum,
$I$ can be approximated by
  \bean
   I &\simeq& \frac{8\pi}{2M+3}\left[ B_\in^{(1)4}\lambda^3\right]
     \left(\la k_d\right)^{2M+3}  \;.
  \eean
Hence the functions $F(k)$, $g(t')$ are given by
  \bean
  &&  F(k)= \frac{\kappa^2}{36(2\pi)^4}
        \frac{1}{2M+3}\left(\la k_d\right)^{2M+3}
        \left[ B_\in^{(1)4}\lambda^3\right] \;,  \\
  &&  g(t')= 1 \;.
  \eean

  In the case $2M+3<0$, we obtain
  \bean
   I &\simeq& 8 \pi\left[ B_\in^{(1)4}\lambda^3\right]
       (\lambda k)^{2M+3}
     \left(\frac{1}{M+3}-\frac{1}{2M+3}\right)  \;.
  \eean
This case is totally coherent and we can set
  \bean
    F(k)&=& \frac{\kappa^2}{36(2\pi)^4}
        \left[ B_\in^{(1)4}\lambda^3\right]\times \\  &&
   \left(\frac{1}{M+3}-\frac{1}{2M+3}\right)(\la k)^{2M+3}   \;,  \\
  g(t') &=& 1 \, .
  \eean

  We now solve Eq.~(\ref{eq.Sigma_2}) for the two different source terms.

\begin{itemize}
\item
  In the case $2M+3>0$, 
we can write
Eq.~(\ref{eq.Sigma_2}) in the form
  \bean
  &&  P''-\frac{4}{z}P'+\left(1+\frac{4}{z^2}\right)P=\frac{\alpha}
       {k^2}~,    \\
  &&  z\equiv kt ~, \qquad |P|\equiv
      \sqrt{\mathcal{P}_{\Sigma}^{(2)\,\Pi}(k,t)}  ~, \\
  &&  \alpha \equiv \sqrt{F(k)}  \;.
  \eean
Solving the above equation on super-horizon scales and
following the considerations for the choice of initial conditions
explained in Appendix~\ref{solution}, we find
  \bean
    P(z)\simeq -
        \frac{\alpha}{2k^{2}}z^2 ~, \qquad z\ll 1 ~,
  \eean
this gives the second order power spectrum
  \bea
    \hspace{+5mm}
    k^3\mathcal{P}_{\Sigma}^{(2)\,\Pi}(k,t) &\simeq &
      \frac{\kappa^2}{36 (2\pi)^4(2M+3)}
      \left[\frac{k^3\PP_{B\,\in}^{(1)}(k)}{H_\in^2}\right]^2 \times
      \nonumber \\ && \hspace{-5mm}
      \left(\frac{a}{a_\in}\right)^4 
      \left(\frac{k_d}{k}\right)^{2M+3}   \;, \quad kt\ll 1~.
  \eea
This is equivalent to a density parameter for $\Sigma^{(2)}$ given by
\bea
   \frac{{\rm d}\Omega_{\rm GW}^{(2)\,\Pi}(k,t)}{{\rm d}\log k} &\simeq&
       \left[\frac{{\rm d}\Omega_{B\,\in}^{(1)}(k)}{{\rm d}\log k} \right]^2
        \left(\frac{k_d}{k}\right)^{2M+3}  
         \nonumber \\ 
         &\simeq& 
         \left[\frac{{\rm d}\Omega_{B\,\in}^{(1)}(k_d)}{{\rm d}\log k}
         \right]^2
         \left(\frac{k}{k_d}\right)^{3}~,   \nonumber \\  \label{O2GWPisup}
     &&      \qquad     \quad kt\ll 1~.
  \eea
Inside the horizon, the Green function oscillates and we can neglect the 
contribution from the source. The solution for the power spectrum is then 
given by
\bea
    &&   \hspace{-6mm} k^3\mathcal{P}_{\Sigma}^{(2)\,\Pi}(k,t) \simeq
      \frac{\kappa^2}{36 (2\pi)^4(2M+3)} \times    \nonumber \\ && 
     \quad \left[\frac{k^3\PP_{B\,\in}^{(1)}(k)}{H_\in^2}\right]^2
      \left(\frac{a}{a_\in}\right)^4 
      \left(\frac{k_d}{k}\right)^{2M+3}   \,,  \nonumber \\
 && \qquad \quad kt\gg 1~.
\eea
Therefore, the  second order density parameter is given by the same expression,
\bea
   \frac{{\rm d}\Omega_{\rm GW}^{(2)\,\Pi}(k,t)}{{\rm d}\log k} &\simeq&
         \left[\frac{{\rm d}\Omega_{B\,\in}^{(1)}(k_d)}{{\rm d}\log k}
         \right]^2
         \left(\frac{k}{k_d}\right)^{3}~,     \nonumber \\  && \qquad
        \quad kt\gg 1~.
  \eea

Up to logarithmic factors this result agrees with the findings of
Ref.~\cite{CD}.

\item
  In the case $2M+3<0$ we have again to solve the equation
  \be
      \label{eq.gw}
  P''-\frac{4}{z}P'+\left(1+\frac{4}{z^2}\right)P=\frac{\alpha}
      {k^2} ~,   
  \ee
Hence
  \bean
    P(z)\simeq -\frac{\alpha}{2k^2}z^2  ~,
    \quad z \ll 1 \,.
  \eean
  But now
\bean
&&  \alpha \equiv \frac{\kappa}{6(2\pi)^2}
        \sqrt{ \frac{1}{2}
        \left(\frac{1}{M+3}-\frac{1}{2M+3}\right)k^{2M+3}}
        \nonumber\\ && \qquad
        \times \left[ B_\in^{(1)2}\lambda^3\right]  \lambda^{M}   \;, \nonumber
\eean
so that
  \bea
    k^3\mathcal{P}_{\Sigma}^{(2)\,\Pi}(k,t) &\simeq&
      \frac{\kappa^2}{ 144 (2\pi)^4}
      \left[\frac{k^3\PP_{B \,\in}^{(1)}(k)}{H_\in^2}\right]^2 \times \nonumber \\
      &&  \left(\frac{a}{a_\in}\right)^4 \,,
      \quad kt \ll 1~.
  \eea
As in the first case, the density parameter is the same for $kt<1$ and $kt>1$,
\be
    \frac{d\Omega_{\rm GW}^{(2)\,\Pi}(k,t)}{d\log k} \simeq
      \left[\frac{d\Omega_{B\,\in}^{(1)}(k)}{d\log k} \right]^2  \,.
\ee
\end{itemize}

\subsubsection{Gravity waves part of the source
$\left[k^3\PP_\Sigma^{(2)\,{\rm GW}}(k,t)\right]$  }
Let us finally consider the part of the source given
by first order gravity waves.
In this case, we can write the source $f_{ij}$ as:
\bea
    && f_{ij}({\bf x},t)=
        \Big[-\frac{3}{2} (1+3w)\HH^2
        {\Sigma_{\langle i}^{(1)}}^{n}
        {\Sigma^{(1)}_{j\rangle n}}+
        \nonumber \\ && \qquad
        6\HH{\Sigma_{\langle i}^{(1)}}^{n}
        \dot{\Sigma}^{(1)}_{j\rangle n}- 3\HH
        {{\dot\Sigma}_{\langle i}^{(1)n} }
        {\Sigma^{(1)}_{j\rangle n}} \Big] \left(\frac{a_\in}{a}\right)^2
        \;.
\eea
  As before, we ignore the traces that are present
  in the above products, once we evaluate them in the momentum space,
since we project them out with $(1/2){\MM_{ij}}^{lm}$ afterwards.
  Remembering that $\Sigma_{ij}=\Sigma^{ij}$,
  we have on super-horizon scales, where the transfer function is given
  by Eq.~(\ref{tranf.func.}):
  \bean
  &&  \left[{{\Sigma_{\langle i(1)}}^{n}}
       {\Sigma^{(1)}_{j\rangle n}}\right]({\bf k},t)=
       \frac{1}{2(2\pi)^3} \left(\frac{a}{a_\in}\right)^8 \times
       \nonumber \\ && \qquad
       {\MM_{ij}}^{lm}
       ({\hat \bk})
       \int {\rm d}^3q \Sigma_{ln}^{(1)\,\in}({\bf q})
       \Sigma_{nm}^{(1)\,\in}({\bk-\bq})    \;,  \\
  &&  \left[{{\Sigma_{\langle i(1)}}^{n}}
       {\dot{\Sigma}^{(1)}_{j\rangle n}}\right]({\bf k},t)=
       \left[{{\dot\Sigma_{\langle i(1)}}^{n}}
       {\Sigma^{(1)}_{j\rangle n}}\right]({\bf k},t)=
       \nonumber \\ && \qquad
       \frac{ 2\HH}{(2\pi)^3}\left(\frac{a}{a_\in}\right)^8
       {\MM_{ij}}^{lm} ({ \hat \bk}) \times
           \nonumber \\ && \qquad
       \int {\rm d}^3q \Sigma_{ln}^{(1)\,\in}({\bf q})
       \Sigma_{nm}^{(1)\,\in}({\bk-\bq})     \;.
\eean
These equations are strictly true only on super-horizon scales
where $\Si \propto 1/a^4$. However, since inside the horizon $\Si$
oscillates and the contribution from the source is negligible,
we can use this approximation.
Setting $w=1/3$ we can finally write the source in the form
\bea
    && f_{ij}({\bf k},t) = \frac{9}{2(2\pi)^3}
          \HH^2 \left(\frac{a}{a_\in}\right)^6 \times
           \nonumber \\ && \qquad
          {\MM_{ij}}^{lm}
          ({\hat \bk})
          \int {\rm d}^3p \Sigma_{ln}^{(1)\,\in}({\bf p})
          \Sigma_{nm}^{(1)\,\in}({\bk-\bp})  \;,
          \nonumber \\
          \label{GWsource}
\eea
  and the two-point correlation function of the source part reads:
  \bean
  && \langle f_{ij} ({\bf k},t')
       f_{rc}^{*}({\bf q},t'') \rangle=
       (2\pi)^3 \delta^3({\bk-\bq}) \MM_{ijrc}({\hat \bk}) \times
       \nonumber \\ && \qquad\qquad \qquad
       h(k,t',t'')
       \;, \\
  && h(k,t',t'')= \frac{1}{8(2\pi)^3} U(t',t'')
       I(k)     \;,\\
  && U(t',t'')=\frac{81}{4}\HH^2(t') \HH^2(t'')
       \left[\frac{a(t')}{a_\in}\right]^6
       \left[\frac{a(t'')}{a_\in}\right]^6   \;,  \\
  && I(k)= \MM_{bdlm}(\hat \bk) \int {\rm d}^3p  \left[ \MM_{lnbf}(\hat \bp)
       \MM_{mndf}(\widehat {\bk-\bp})+ \right.
       \nonumber \\ && \qquad \left.
       \MM_{lndf}(\hat \bp)
       \MM_{mndf}(\widehat {\bk-\bp}) \right]
       \PP_{\Sigma\,\in}^{(1)}(p) \PP_{\Sigma\,\in}^{(1)}(|{\bk-\bp}|)    \;.
  \eean
More details about the computation of $h(k,t't'')$ and of the four point
correlation function of the gravity
waves can be found in Appendix \ref{GWcorrelator}.\\
Using the tensor calculus package  ``xAct''  for Mathematica
\cite{xAct}, we can compute the above
products of   the three projectors,
  \bea
  &&  \MM_{bdlm}(\hat \bk) \left[ \MM_{lnbf}(\hat \bp)
       \MM_{mndf}(\widehat {\bk-\bp})+
       \right.
       \nonumber \\ && \qquad \left.
       \MM_{lndf}(\hat \bp)
       \MM_{mnbf}(\widehat {\bk-\bp}) \right] =
       \nonumber \\ && \qquad
       2(1+\alpha^2+\beta^2+\alpha^2\beta^2-  8\alpha\beta\gamma +
       \nonumber \\ && \qquad
       \gamma^2+\alpha^2\gamma^2+\beta^2\gamma^2+
       \alpha^2\beta^2\gamma^2)
       \simeq 2\;.
       \nonumber \\
      \label{XtensorProduct}
  \eea
  where $\alpha\equiv \hat{k}\cdot (\widehat{k-p})$,
  $\beta\equiv \hat{p}\cdot (\widehat{k-p})$
  and   $\gamma\equiv \hat{k}\cdot \hat{p}$. Again we have approximated this
 angular dependence by a constant to simplify the calculations. This
approximation is well justified within our accuracy.
  In order to write the function $h(k,t',t'')\simeq
  F(k)g(t')g(t'')$, we have to evaluate the integral $I$ as
  before. We first consider the most interesting case of a scale invariant
spectrum, {$A\simeq-3$}. Up to an infrared log-divergence which we neglect
as usual (this divergence can be avoided if we choose $A=-2.99$ instead of
$A=-3$), we have
  \bean
  &&  F(k)\simeq \frac{81\pi}{2(2\pi)^3}\left[k^3\PP_{\Sigma\,\in}^{(1)}(k)\right]^2
            \frac{1}{k^3}
           \;, \\
  &&  g(t') \simeq \HH^2(t') \left[\frac{a(t')}{a_\in}\right]^6
           \;.
  \eean
  Therefore,  the equation for
  $\sqrt{\PP_{\Sigma}^{(2)\,\rm GW}(k,t)}$ in the radiation dominated
  era becomes
  \bean
  &&  P''(z)-\frac{4}{z}P'+\left(1+\frac{4}{z^2}\right)P=\frac{\alpha}{k^6}z^4
           ~, \\
  &&  z \equiv kt ~, \qquad |P|\equiv \sqrt{\PP_{\Sigma}^{(2)\,\rm
          GW}(k,t)} ~,\\
  &&  \alpha \equiv \frac{1}{t_\in^6}\sqrt{\frac{81\pi}{2(2\pi)^3}
             \frac{\left[k^3\PP_{\Sigma\,\in}^{(1)}(k)\right]^2}{k^3}}  ~.
  \eean
  The super-horizon solution,
  evaluated always with the help of the Wronskian method
  and keeping only the non-homogeneous part as explained in the Appendix
  \ref{solution},
  is the given by
  \bean
    P(z)\simeq \frac{1}{10}\frac{\alpha}{k^6}z^6 ~, \qquad z \ll 1 ~,
  \eean
  that yields a contribution to the gravity wave power spectrum given by
  \bea
    k^3\mathcal{P}_{\Sigma}^{(2)\,\rm GW}(k,t) &\simeq&
      0.4\frac{\pi}{(2\pi)^3}
      \left[k^3\PP_{\Sigma\,\in}^{(1)}(k)\right]^2
      \left(\frac{a}{a_\in}\right)^{12} \,,
        \nonumber \\ && kt \ll 1~.
  \eea
  For the density parameter on super-horizon scales this yields
  \bea
     \frac{{\rm d}\Omega_{\rm GW}^{(2)\,\Si}(k,t)}{{\rm d}\log k} &\simeq&
       0.01
       \left[\frac{{\rm d}\Omega_{\rm GW}^{(1)}(k,t)}{{\rm d}\log k} \right]^2
       \,, 
       \nonumber \\ &&  kt \ll 1 \,.
  \eea
  Considering now the sub-horizon limit, we obtain for the
  power spectrum the following expression:
  \bea
  &&  k^3\mathcal{P}_{\Sigma}^{(2)\,\rm GW}(k,t) \simeq
       0.1\frac{\pi}{(2\pi)^3}
       \left[k^3\PP_{\Sigma\,\in}^{(1)}(k)\right]^2\times 
       \nonumber \\  && \qquad \quad
       \frac{1}{(kt_\in)^{8}}
       \left(\frac{a}{a_{\rm in}}\right)^4 ~, \quad kt \gg 1 ~,
  \eea
  and the density parameter becomes
  \bea
     \frac{{\rm d}\Omega_{\rm GW}^{(2)\,\Si}(k,t)}{{\rm d}\log k} \simeq
       0.02
       \left[\frac{{\rm d}\Omega_{\rm GW}^{(1)}(k)}{{\rm d}\log k} \right]^2
       ~, \quad kt \gg 1 \,.
       \nonumber \\
  \eea

  On the other hand, when {$2A+3>0$} we have
  \bean
  &&  F(k)\simeq\frac{81\pi}{2(2\pi)^3}\Sigma_{(1)\,\in}^4 \lambda^{6+2A}
        \frac{1}{2A+3}
        \;, \\
  &&  g(t') \simeq \HH^2(t') \left[\frac{a(t')}{a_\in}\right]^6
        \left(\frac{1}{t'}\right)^{(2A+3)/2}    \;.
   \eean
  In the radiation epoch the equation for
  $\sqrt{\PP_{\Sigma}^{(2)\,\rm GW}(k,t)}$   reads:
  \bean
  &&  {P''}-\frac{4}{z}P'+\left(1+\frac{4}{z^2}\right)P=
        \alpha
         z^{(5/2-A)}   \;,  \\
  && \alpha \equiv
        \sqrt{\frac{81\pi}{2(2\pi)^3}
        \frac{1}{2A+3} }k^{3/2}\PP_{\Sigma\,\in}^{(1)}(k)\frac{1}{(kt_\in)^6} ~.
  \eean
  Solving the above equation in the long wavelengths limit, we find:
  \bean
    P(z)\simeq \frac{\alpha}{2}
      z^{9/2-A}    \,, \qquad z \ll 1~,
  \eean
  where the exact pre-factor depends weakly on the value of $A$.
  For the power spectrum this results in
  \bea
  k^3\mathcal{P}_{\Sigma}^{(2)\,\rm GW}(k,t) &\simeq&
       \frac{81\pi}{8(2\pi)^3} \left[k^3 \PP^{(1)}_{\Si\,\in}(k) \right]^2
       \left(\frac{a}{a_\in}\right)^{12} \times 
       \nonumber \\ && 
       (kt)^{-2A-3}~,   \qquad kt \ll 1 \,.
  \eea
and inside the horizon this reads
  \bea
  k^3\mathcal{P}_{\Sigma}^{(2)\,\rm GW}(k,t) &\simeq&
       \frac{81\pi}{16(2\pi)^3} \left[k^3 \PP^{(1)}_{\Si\,\in}(k) \right]^2
       \left(\frac{a}{a_\in}\right)^{4} \times 
       \nonumber \\ && 
       \frac{1}{(kt_\in)^8}~,   \qquad kt \gg 1 \,.
  \eea
Translating this to the density parameter as above, we obtain
\bea
  \frac{{\rm d}\Omega_{\rm GW}^{(2)\,\Si}(k,t)}{{\rm d}\log k} &\simeq&
       0.2
       \left[\frac{{\rm d}\Omega_{\rm GW}^{(1)}(k)}{{\rm d}\log k} 
       \right]^2(kt)^{-2A-3}
       ~, \nonumber \\ 
   &\simeq&   0.2
       \left[\left.\frac{{\rm d}\Omega_{\rm GW}^{(1)}(k)}{{\rm d}\log k} 
       \right|_{k=1/t}\right]^2(kt)^{3}
       ~, \nonumber \\
&& kt \ll 1 \,,
       \nonumber \\
  \frac{{\rm d}\Omega_{\rm GW}^{(2)\,\Si}(k,t)}{{\rm d}\log k} &\simeq&
       0.1
       \left[\frac{{\rm d}\Omega_{\rm GW}^{(1)}(k)}{{\rm d}\log k} \right]^2
       ~, \quad kt \gg 1 \,.
        \nonumber \\
  \eea

\section{Summary and conclusions}
In this work we have studied the evolution of stochastic cosmic magnetic
fields and gravity waves up to second order in the perturbations. We have
especially calculated the density parameters of the generated second order
perturbations. We start with density parameters
$ \left[{\rm d}\Omega_{\rm B}^{(1)}(k,t)/{\rm d}\log k\right]$ and
$\left[{\rm d}\Omega_{\rm GW}^{(1)}(k,t)/{\rm d}\log k\right]$ 
which are related to the first order magnetic field and gravitational wave 
power spectra in Section~\ref{s:O1}. Since tensor
perturbations  grow on super-horizon scales, the gravity wave density
parameter grows on super-Hubble scales and only becomes constant once the
perturbations enter the horizon. For perturbation theory to be valid, we
have of course to require that these density parameters are much smaller
than unity. As we have seen in Section~\ref{s:O1}, to require that
$\left[{\rm d}\Omega_{\rm GW}^{(1)}(k,t)/{\rm d}\log k\right]$ 
is smaller than one also on
sub-Hubble scales, is equivalent to
\be
\frac{{\rm d}\Omega_{\rm GW}^{(1)}(k,t_\in)}{{\rm d}\log k}
  \frac{1}{(kt_\in)^4} 
  \simeq \left( \frac{H_{\rm inf}}{M_{\rm P}} \right)^2  \ll 1 \; .
\ee

Here we summarize the new results on the density parameters for
second order perturbation on sub-horizon scales. For magnetic fields,
 we obtain
\be
\frac{{\rm d}\Omega_{\rm B}^{(2)}(k)}{{\rm d}\log k} \simeq
\frac{{\rm d}\Omega_{\rm GW}^{(1)}(k)}{{\rm d}\log k}
\frac{{\rm d}\Omega_{\rm B}^{(1)}(k)}{{\rm d}\log k}\,, \quad  tk\gg 1 ~,
\ee
up to numerical constants and logarithms which are beyond the accuracy of 
our approximation.
Hence, it is not correct that the presence of gravity waves resonantly enhances
a first order magnetic field. The second order density parameter is quite
what we would naively expect and it is much smaller than the first order
perturbations as long as the latter are small.
Also on super-horizon scales, the second order magnetic field density
parameter is always much smaller than the first order one, see 
Eqs.(\ref{e:B2red}) and (\ref{e:B2blue}).

Since the growth comes from super horizon scales, conductivity is not 
relevant for this result. We have shown that also in a matter 
dominated background we obtain 
\bea
\frac{{\rm d}\Omega_{\rm B}^{(2)}(k)}{{\rm d}\log k} &\simeq& 
\left.\frac{{\rm d}\Omega_{\rm GW}^{(1)}(k)}{{\rm d}\log k}\right|_{kt=1}
\frac{{\rm d}\Omega_{\rm B}^{(1)}(k)}{{\rm d}\log k} \nonumber\\
&\simeq& \left(\frac{H_{\rm{inf}}}{M_{\rm P}}\right)^2
\frac{{\rm d}\Omega_{\rm B}^{(1)}(k)}{{\rm d}\log k} \\
&\ll& \frac{{\rm d}\Omega_{\rm B}^{(1)}(k)}{{\rm d}\log k}~,
\eea
hence no significant amplification.

Second order gravity waves are induced on the one hand by the anisotropic
stresses of the first order magnetic fields and on the other hand by the
quadratic terms in the evolution equation for $\si_{ij}$ which are, e.g., of
the form $\si_{im}{\dot\si^m}_j$ and similar expressions. In
Section~\ref{s:O2Pi} we have shown that the second order contribution from
anisotropic stresses on sub-Hubble scales is of the order of
\be
 \frac{{\rm d}\Omega_{\rm GW}^{(2)\,\Pi}(k,t)}{{\rm d}\log k} \simeq 
\left\{ \begin{array}{ll}
      \left[\frac{{\rm d}\Omega_{B}^{(1)}(k_d)}{{\rm d}\log k_d}
        \right]^2\left(\frac{k}{k_d}\right)^3 , & 
\mbox{if } 2M+3>0 \\   & \\
\left[\frac{{\rm d}\Omega_{B}^{(1)}(k)}{{\rm d}\log k}\right]^2 , & 
\mbox{if } 2M+3<0 ~, \end{array} \right.
\ee
both on super- and sub-horizon scales.
Note that the above expression is continuous at $2M+3=0$, where both
expressions scale like $(k\la)^3$ and are independent of $k_d$.
One should point out that we neglected the slow time dependence
of the damping scale. Correctly one has to choose the value of the 
damping scale at horizon crossing, $k_d(t_k)$ with $t_k=1/k$.
Depending on the magnetic field spectrum, the resulting gravity waves
come mainly from the small scale magnetic field, if its spectrum is blue
$2M+3>0$.
In this case the gravity waves power spectrum is always
proportional to $k^3$. In our case of a simple power law magnetic field 
spectrum, this behavior is maintained for all $k<k_d$.
If the magnetic field spectrum is red, $2M+3<0$, gravity waves depend on
the field at scale $k$ and their spectrum is the square of the $B$-field 
spectrum.
In the first case, the non-linearity leads to a 'sweeping' of magnetic 
field power on small scales to gravitational wave power on larger scales. 
This can be regarded as an 'inverse cascade' of small scale magnetic field 
power into large scale gravity waves.
But in no case can the gravity wave density parameter become larger
than the one of the magnetic field, which has to be much smaller than one, for
perturbation theory to be valid.

A similar result was already obtained in Ref.~\cite{CD}. Contrary to this 
reference we have no logarithmic build-up of gravity waves. This comes from our
different treatment; we directly calculate the shear $\si_{ij}$ and not the
tensor perturbation of the metric, $h_{ij}$. In this way we loose the log
term which corresponds to the homogeneous $h_{ij}=$ constant solution on
super-horizon scales to which we are not sensitive. However, in our more
qualitative work, we do not want to insist on log terms which we neglect in
this work also in other places.

The second order
gravity wave density parameter induced by first order gravity waves
is given by
\be
  \frac{{\rm d}\Omega_{\rm GW}^{(2)\,\Si}(k,t)}{{\rm d}\log k} \simeq
        \left[\frac{{\rm d}\Omega_{\rm GW}^{(1)}(k,t)}{{\rm d}\log k} \right]^2 ,
        \qquad kt \gg 1 ~ ,
\ee
on sub-horizon scales.\\
Adding both contributions we find
\be
 \frac{{\rm d}\Omega_{\rm GW}^{(2)}(k,t)}{{\rm d}\log k} \simeq 
\left\{ \begin{array}{ll}  &
      \left[\frac{{\rm d}\Omega_{B}^{(1)}(k_d)}{{\rm d}\log k_d}
        \right]^2\left(\frac{k}{k_d}\right)^3 
      + \left[\frac{{\rm d}\Omega_{\rm GW}^{(1)}(k,t)}{{\rm d}\log k} 
        \right]^2  , \\   &
      \qquad \mbox{if } 2M+3>0 \,, \quad kt\gg 1 \\   & \\ &
      \left[\frac{{\rm d}\Omega_{B}^{(1)}(k)}{{\rm d}\log k}\right]^2 
      +\left[\frac{{\rm d}\Omega_{\rm GW}^{(1)}(k,t)}{{\rm d}\log k} 
        \right]^2 , \\ & 
      \qquad \mbox{if } 2M+3<0  \,, \quad kt\gg 1 ~. \end{array} \right.
\ee

\section*{Acknowledgments}

We are grateful to Roy Maartens, Christos Tsagas, Chiara Caprini and 
Cyril Pitrou for helpful
discussions. EF thanks the ``EARA Early Stage Training'' fellowship
for financial support.
We acknowledge support by the Swiss National Science Foundation.

\vspace{1.3cm}

\appendix

\section{General Solution of a Differential Equation with the
  Wronskian Method}
\label{solution}

Here we discuss in detail the Wronskian method with which we find the
solution of the differential equations in this paper.
If we have a inhomogeneous linear second order equation with inhomogeneity
$S(z)$, its most general solution is of the form
$$ P(z)= c_1(z)P_1(z) + c_2(z)P_2(z) + a_1P_1(z) + a_2P_2(z) \, ,$$
where $P_1(z)$ and $P_2(z)$ are two (linearly independent)
homogeneous solutions which we suppose
to be known, $W(z)=P_1P_2'-P_1'P_2$ is their Wronskian, and
\bean
 c_1(z) &=& -\int_{z_\in}^{z} dx \frac{S(x)}{W(x)}P_2(x)   ~,\\
  c_2(z)&=& \int_{z_\in}^{z} dx \frac{S(x)}{W(x)}P_1(x)   ~.\\
\eean
The particular solution given by the first two terms is such that
$P_{\rm inh}(z)= c_1(z)P_1(z) + c_2(z)P_2(z)$ vanishes at $z=z_\in$ and also
$P'_{\rm inh}(z_\in)=0$. The general solution is obtained by adding a
 homogeneous solution, $P_{\rm hom}(z) = a_1P_1(z) + a_2P_2(z)$ with
arbitrary constants $a_1$ and $a_2$.

Let us first consider the example given in Eq.~(\ref{eq.gw}),
\bean
  P''-\frac{4}{z}P'+\left(1+\frac{4}{z^2}\right)P=\frac{\alpha}{k^2}  ~,
\eean
where $\al/k^2$ is a constant source term.
The homogeneous solutions are given by $P_1(z)=z^3 j_1(z)$ and
$P_2(z)=z^3 y_1(z)$ and the Wronskian determinant reads
\bean
  W(z)=z^3   ~.
\eean
In the regime $z\ll 1$ we can approximate the spherical Bessel functions by
powers and we find the following general expression for $P(z)$:
\be     \label{general solution}
  P(z)=-\frac{\alpha}{3k^2}\left(\frac{z^2}{2}\!-\frac{z^4}{2z^2_\in}
           \!+z^2 \! - zz_\in\right)  +a_1z^4\! + a_2z  ~.
\ee
where we have used the fact that, when $z \ll 1$, we can approximate
$P_1(z)\simeq z^4$ and $P_2(z)\simeq -z$.
Now it is important to notice that the second and the fourth terms
of the inhomogeneous solution
(\ref{general solution}) have the same functional behavior as
homogeneous solutions and we can always choose $a_1$ and $a_2$
such that the homogeneous part cancels them. This is actually always
true for the  contributions from the lower boundary of the
inhomogeneous solution. This may sound pedantic, but it is very important
in this specific case as the second term in (\ref{general solution}) dominates
if it is present. In our analysis we have always subtracted such ``homogeneous
contributions'' and only kept the ``minimal part'', which in this case is
\bea
  P(z)\simeq -\frac{\alpha}{2k^2}z^2  ~, \quad z \ll 1 \,.
\eea
This procedure is important and it is responsible for the results which
we have obtained. We justify it also by the fact that the first order solution
has exactly the the same time evolution as the homogeneous term and therefore a
term $\propto z^4$ present at early times, should be included in the first
order perturbations. Once the wave number has entered the horizon, $z\gg 1$,
the Green function starts to oscillate and the additional contribution to the
integral can be neglected. We then can match the inhomogeneous solution at 
horizon crossing to the homogenous one at later times. Up to matching 
details which we have not considered, this yields
\be
 P(z)\simeq \frac{\alpha}{2k^2}z^2\cos z ~, \quad z\gg 1  ~.
\ee

In the same way, we deal with Eq.~(\ref{eq.B})
\be
  P''-\frac{4}{z}P' +\left(1+\frac{4}{z^2}\right)+P=\frac{\alpha}{k^2}  ~.
\ee
 The homogeneous
solutions are $P_1(z)=z^2 j_0(z)\simeq z^2$ and
$P_2(z)=z^2 y_0(z)\simeq -z$. These approximations are valid for $z\ll 1$.
Using again the Wronskian method, we
obtain the following general solution on super-Hubble scales, $z\ll 1$:
\bea
  P(z) &=& \frac{\alpha}{k^2}\left(z^2\log\left(\frac{z}{z_\in}\right)
     -z^2+zz_\in \right)  +a_1z^2 + a_2 z~,  \nonumber \\
   &&  \quad z\ll 1 ~,
\eea

Here, the homogeneous solution parts are $-z^2$ and $zz_\in$,
therefore we can identify the solution due to the presence of the source
again as
\be
  P(z)\simeq \frac{\alpha}{k^2}z^2\log\left(\frac{z}{z_\in}\right)  
    \,, \quad  z\ll 1 ~.
\ee
On sub-horizon scales this becomes, up to matching details which only modify
the phase and have an irrelevant effect on the pre-factors,
\be
  P(z)\simeq \frac{\alpha}{k^2}\log(kt_\in)z\cos z  \,, \quad  z\gg 1 ~.
\ee
If the source term depends on $z$, the details of the calculation as well as
the results change somewhat, but the basic argumentation remains the same.
We therefore do not repeat the $z$-dependent examples which arise in this
work here.

\section{The four-point correlator of gravity waves}
\label{GWcorrelator}

Starting from Eq.~(\ref{GWsource}), we compute the two-point
correlation function of the source term $\langle f_{ij} ({\bf k},t')
f_{rn}^{*}({\bf p},t'') \rangle$, which is given by
\bea
&& \langle f_{ij} ({\bf k},t')f_{rc}^{*}({\bf q},t'') \rangle =
    \frac{1}{(2\pi)^6} U(t',t'') {\MM_{ij}}^{lm}(\hat \bk) \times
    \nonumber \\ && \quad
    {\MM_{rc}}^{bd}(\hat \bq)
    \int {\rm d}^3p \int {\rm d}^3s \langle \Si^{(1)\,\in}_{ln}(\bp)
    \Si^{(1)\,\in}_{nm}(\bk-\bp) \times
    \nonumber \\ && \quad
    \Si^{*(1)\,\in}_{bf}({\bf s})
    \Si^{*(1)\,\in}_{fd}({\bf q-s}) \rangle   \,,
\label{2pGWsource}
\eea
where the function $U(t',t'')$ contains the all time-dependence of the above
expression:
\bea
  U(t',t'')= \frac{81}{4}\HH^2(t')\HH^2(t'')\left[\frac{a(t')}{a_\in}\right]^6
      \left[\frac{a(t'')}{a_\in}\right]^6    \,.
\eea

To compute the four-point correlator, we assume that the
random variables that describe  gravity waves are Gaussian, therefore
we can apply Wick's theorem. The we can write the products of
four gravity waves $\Si^{(1)}$ as
\bea
&&\hspace{-4mm}
 \langle \Si^{(1)\,\in}_{ln}(\bp)
    \Si^{(1)\,\in}_{nm}(\bk-\bp)\Si^{*(1)\,\in}_{bf }({\bf s})
    \Si^{*(1)\,\in}_{fd}({\bf q}-{\bf s}) \rangle =
    \nonumber \\ && \;
    \langle \Si^{(1)\,\in}_{ln}(\bp)  \Si^{*(1)\,\in}_{bf}({\bf s})
    \rangle \langle
    \Si^{(1)\,\in}_{nm}(\bk-\bp)
    \Si^{*(1)\,\in}_{fd}({\bf q-s}) \rangle +
    \nonumber \\ && \;
    \langle \Si^{(1)\,\in}_{ln}(\bp)  \Si^{*(1)\,\in}_{fd}({\bf q}-{\bf s})
    \rangle \langle
    \Si^{(1)\,\in}_{nm}(\bk-\bp)  \Si^{*(1)\,\in}_{bf}({\bf s})
     \rangle +
     \nonumber \\ &&  \;
     \langle \Si^{(1)\,\in}_{ln}(\bp)
    \Si^{(1)\,\in}_{nm}(\bk-\bp) \rangle \langle
    \Si^{*(1)\,\in}_{bf }({\bf s})
    \Si^{*(1)\,\in}_{fd}({\bf q}-{\bf s}) \rangle   \,.
    \nonumber \\
\label{Wick}
\eea
Once the double integration is performed, the last term contributes a
constant $\propto \de^3(\bk)$ which can be disregarded (a background term).
Integrating the remaining  two terms
over $d^3s$, we can eliminate one of the two $\delta$-functions which
come from the expression of the two point gravity wave correlator.
Using the reality condition, $\Si^{*}_{ij }({\bk}) =
\Si_{ij }(-\bk)$, and the expression for the two-point correlation function
of gravity waves given in Eq.~(\ref{2pGW}), we then obtain
\bea
&&  \int {\rm d}^3p \int {\rm d}^3s \langle \Si^{(1)\,\in}_{ln}(\bp)
      \Si^{(1)\,\in}_{nm}(\bk-\bp)\Si^{*(1)\,\in}_{bf}({\bf s}) \times
      \nonumber \\ && \quad
      \Si^{*(1)\,\in}_{fd}({\bf q}-{\bf s}) \rangle =
      (2\pi)^6\delta^3(\bk-\bq)
      \int {\rm d}^3p \PP_{\Si\,\in}^{(1)}(p)  \times
      \nonumber \\ && \quad
      \PP_{\Si\,\in}^{(1)}(|\bk-\bp|)
      \left[\MM_{lnbf}(\hat \bp)\MM_{mndf}(\widehat{\bk-\bp})
      \right.\nonumber \\ && \quad \left.
      + \MM_{lndf}(\hat \bp)\MM_{mnbf}(\widehat{\bk-\bp}) \right]
         \,.
\eea
The above equation is symmetric in
$\bk$ and $\bq$, as well as under the exchange of the first and second
pairs of indices. Moreover, it is symmetric under the exchange
of the first index with the second and the third with the fourth.
This suggests us to write the two point correlation function
of the source term as
\bea
  \langle f_{ij} ({\bf k},t')f_{rc}^{*}({\bf q},t'') \rangle &=&
    (2\pi)^3 \delta^3(\bk-\bq) \MM_{ijrc}(\hat \bk) \times
    \nonumber \\  &&
    h(k,t',t'')  \,,
\eea
since the tensor $\MM_{ijrc}$ has the same symmetries.

To obtain an expression for the function $h(k,t',t'')$, it is sufficient to
calculate the trace of the above two point correlator.
We hence should multiply the r.h.s. of the above equation and of
Eq.~(\ref{2pGWsource}) by  $\MM^{ijrc}(\hat \bk)$.
Then, setting them to be equal and remembering
that $\MM^{ijrc}\MM_{ijrc}=8$ \cite{CDK}, we obtain
\bea
&& \hspace{-4mm}
  8(2\pi)^3 \delta^3(\bk-\bq) h(k,t',t'')=
     U(t',t'') \delta^3(\bk-\bq) \times
     \nonumber \\ && 
     \MM^{crlm}(\hat \bk){\MM^{bd}}_{rc}(\hat \bq)
     \int {\rm d}^3p \left[\MM_{lnbf}(\hat \bp)\MM_{mndf}(\widehat{\bk-\bp})+
       \right.\nonumber \\ &&  \left.
       \MM_{lndf}(\hat \bp)\MM_{mnbf}(\widehat{\bk-\bp}) \right]
      \PP_{\Si\,\in}^{(1)}(p) \PP_{\Si\,\in}^{(1)}(|\bk-\bp|) 
      \nonumber \\
\eea
with
\bea
&& \hspace{-5mm} 
  h(k,t',t'')= \frac{1}{8(2\pi)^3} U(t',t'') \MM_{bdlm}(\hat \bk)
      \times
      \nonumber \\ && \hspace{-3mm}
      \int {\rm d}^3p 
      \PP_{\Si\,\in}^{(1)}(p) \PP_{\Si\,\in}^{(1)}(|\bk-\bp|) \times
      \nonumber \\ && \hspace{-3mm}
      \left[\MM_{lnbf}(\hat \bp)\MM_{mndf}(\widehat{\bk-\bp})
        + \MM_{lndf}(\hat \bp)\MM_{mnbf}(\widehat{\bk-\bp}) \right] \,.
      \nonumber \\
\eea

Finally, we have to perform the above product of three polarization
tensors, defined as in Eq.~(\ref{e:MM}). To achieve this aim, we use the
free source package ``xAct'' for Mathematica \cite{xAct}: it is sufficient to
define
a three dimensional flat metric and the projection tensor
$\PP_{ij}(\hat\bk)=\de_{ij}-k^{-2}k_ik_j$ onto the plane normal to $\bk$.
Then, we can express $\MM_{ijlm}(\hat \bk)$ in terms of this projector as
\bean
  \mathcal{M}_{ijlm} \equiv \mathcal{P}_{il}\mathcal{P}_{jm} +
      \mathcal{P}_{im}\mathcal{P}_{jl}
      - \mathcal{P}_{ij} \mathcal{P}_{lm}   \,.
\eean
Defining the angles between the three directions as
$\alpha\equiv \hat{k}\cdot (\widehat{k-p})$,
$\beta\equiv \hat{p}\cdot (\widehat{k-p})$
and   $\gamma\equiv \hat{k}\cdot \hat{p}$, we obtain the expression given
in Eq.~(\ref{XtensorProduct}).


\end{document}